\documentclass[conference,letterpaper]{IEEEtran}

\usepackage[utf8]{inputenc} 
\usepackage[T1]{fontenc}
\usepackage{url}
\usepackage{ifthen}
\usepackage{cite}
\usepackage[cmex10]{amsmath} 
\usepackage{mathtools}
\usepackage{amsfonts}
\usepackage{pgfplots}
\pgfplotsset{compat=newest}
\usepgfplotslibrary{groupplots}
\usetikzlibrary{matrix}
\usepgfplotslibrary{dateplot}
\newcounter{plot}[figure]

\usepackage{stmaryrd, amssymb}
\usepackage{tikz}
\usetikzlibrary{positioning}
\usepackage{cleveref}
\crefname{plot}{plot}{plots}
\Crefname{plot}{Plot}{Plots}

\usepackage{algorithm}
\usepackage[noend]{algpseudocode}
\algrenewcommand{\algorithmiccomment}[1]{\hskip3em // #1}
\algrenewcommand\algorithmicindent{0.8em}%
\usepackage{etoolbox}

\makeatletter
\newcommand*{\algrule}[1][\algorithmicindent]{%
  \makebox[#1][l]{%
    \hspace*{.2em}
    \vrule height .75\baselineskip depth .25\baselineskip
  }
}

\newcount\ALG@printindent@tempcnta
\def\ALG@printindent{%
    \ifnum \theALG@nested>0
    \ifx\ALG@text\ALG@x@notext
    \else
    \unskip
    \ALG@printindent@tempcnta=1
    \loop
    \algrule[\csname ALG@ind@\the\ALG@printindent@tempcnta\endcsname]%
    \advance \ALG@printindent@tempcnta 1
    \ifnum \ALG@printindent@tempcnta<\numexpr\theALG@nested+1\relax
    \repeat
    \fi
    \fi
}
\patchcmd{\ALG@doentity}{\noindent\hskip\ALG@tlm}{\ALG@printindent}{}{\errmessage{failed to patch}}
\patchcmd{\ALG@doentity}{\item[]\nointerlineskip}{}{}{} 
\makeatother

\usepackage{flushend}

\usepackage{acronym}
\acrodef{OSD}{ordered-statistics decoding}
\acrodef{LDPC}{low-density parity-check}
\acrodef{QLDPC}{quantum low-density parity-check}
\acrodef{BP}{belief propagation}
\acrodef{BPGD}{belief propagation guided decimation}
\acrodef{BPGDG}{BPGD guessing}
\acrodef{GDG}{guided decimation guessing}
\acrodef{VN}{variable node}
\acrodef{CN}{check node}
\acrodef{PCM}{parity-check matrix}
\acrodefplural{PCM}{parity-check matrices}
\acrodef{LLR}{log-likelihood ratio}
\acrodef{BB}{bivariate bicycle}
\acrodef{SM}{syndrome measurement}
\acrodef{BP+OSD}{belief propagation and ordered-statistics decoding} 
\newcommand{\bA}{\mathbf{A}}
\newcommand{\bB}{\mathbf{B}}

\newcommand{\F}{\mathbb{F}}

\newcommand{\bx}{\mathbf{x}}

\newcommand{\bs}{\mathbf{s}}
\newcommand{\be}{\mathbf{e}}

\newcommand{\bL}{\mathbf{L}}
\newcommand{\bH}{\mathbf{H}}

\newcommand{\hbe}{\hat{\mathbf{e}}}
\newcommand{\hbx}{\hat{\mathbf{x}}}

\newcommand{\ra}{\rangle}

\newcommand{\PM}{\text{PM}}

\newcommand{\wt}{\text{wt}}

\renewcommand{\epsilon}{\ensuremath\varepsilon}

\renewcommand{\phi}{\ensuremath{\varphi}}

\begin{document}
\title{Toward Low-latency Iterative Decoding of QLDPC Codes Under Circuit-Level Noise} 

\author{%
  \IEEEauthorblockN{Anqi Gong\IEEEauthorrefmark{1},
                    Sebastian Cammerer\IEEEauthorrefmark{2},
                    and Joseph M. Renes\IEEEauthorrefmark{1}}
  \IEEEauthorblockA{\IEEEauthorrefmark{1}%
                    ETH Z\"urich, Switzerland,
                    gonga@student.ethz.ch}
  \IEEEauthorblockA{\IEEEauthorrefmark{2}%
                    NVIDIA}
  \\[-6.0ex]
}

\maketitle

\begin{abstract}
    We introduce a sliding window decoder based on belief propagation (BP) with guided decimation for the purposes of decoding quantum low-density parity-check codes in the presence of circuit-level noise. 
    Windowed decoding keeps the decoding complexity reasonable when, as is typically the case, repeated rounds of syndrome extraction are required to decode.
    Within each window, we employ several rounds of BP with decimation of the variable node that we expect to be the most likely to flip in each round, 
    Furthermore, we employ ensemble decoding to keep both decimation options (guesses) open in a small number of chosen rounds. 
    We term the resulting decoder BP with guided decimation guessing (GDG). 
    Applied to bivariate bicycle codes, GDG achieves a similar logical error rate as BP with an additional OSD post-processing stage (BP+OSD) and combination-sweep of order 10.
    For a window size of three syndrome cycles, a multi-threaded CPU implementation of GDG achieves a worst-case decoding latency of 3ms per window for the [[144,12,12]] code. The source code of this work is available online.\footnote{https://github.com/gongaa/SlidingWindowDecoder}
\end{abstract}

\section{Introduction}
The recent progress in asymptotically good \ac{QLDPC} codes
\cite{almost_linear_distance, asmptotically_good_panteleev, balanced_product, quantum_tanner_codes} renders them a promising candidate for low-overhead fault-tolerant quantum computing. Besides large blocklength codes, numerous short to medium blocklength ($N\lesssim 1000$) \ac{QLDPC} codes with distance near---or even exceeding---the square root of blocklength have been proposed \cite{panteleev2021degenerate,2BGA,bravyi2023highthreshold}. Their parity-check matrices are carefully designed, and some of them have certain structures that are promising for future use \cite{xu2023constantoverhead,viszlai2023matching,Bluvstein_2023}. In particular, the \ac{BB} codes \cite{bravyi2023highthreshold} are numerically shown to be more resource-efficient than the planar surface codes \cite{topological}. 

Due to the noise in the \ac{SM} operations, the \ac{SM} circuit is typically repeated for multiple cycles. 
This comes at the price of a considerable increase in decoding complexity. In this work, we employ a sliding window decoder based on \ac{BPGD} \cite{yao2023BPGD} to handle streaming SM input. 

The basic idea of window decoding is to use the syndrome outputs of a small number of subsequent rounds (the window) to determine the location of faults in the early part of the window, and then slide the window forward by a few rounds and repeat the process. 
An inner decoder is used on each window which ideally finds the lowest-weight correction such that the overall performance degradation compared to global decoding remains tolerable. 
The advantage of this approach is a lower decoding latency; no need to wait for a lengthy decoding process that only begins after collecting the entire set of syndrome data.

Sliding window decoders have been applied to surface codes under circuit-level noise \cite{Alibaba,skoric2023parallel,topological}, where both the SM circuit and measurements are assumed to be noisy. 
Recently, it has also been applied to some QLDPC codes under phenomenological noise \cite{huang2023window} where only the measurements are assumed to be noisy. 
In \cite{huang2023window}, \ac{BP+OSD} \cite{panteleev2021degenerate,roffe2020osd-cs} is used on each window to achieve the aforementioned low-weight requirement.

We propose to use an extended version of the \ac{BPGD} \cite{yao2023BPGD} algorithm as the inner decoder. 
\ac{BPGD} interleaves BP runs with a \emph{global} \ac{VN} selection and subsequent decimation of that VN.
Compared to \ac{BP+OSD}, the advantage is a possibly lower worst-case runtime,  as BPGD avoids the need for Gaussian elimination.
Furthermore, \ac{BPGD} has demonstrated performance comparable to \ac{BP+OSD} on data qubit noise decoding \cite{yao2023BPGD}. 

When applied to decoding circuit-level noise, however, we find that the BPGD \ac{VN} selection rule, which is to choose the most reliable \ac{VN}, becomes less efficient, since in this setting there are significantly more possible fault locations than actual fault occurrences. This motivates us to choose the most likely VN to flip at each step, with the idea that this will lead to BP convergence in fewer steps. 
Moreover, we use a short \emph{history} of posterior \acp{LLR} from recent BP iterations to choose the decimation value for the selected VN. 
Thereby, we build up a main decimation path of low depth.
Lastly, we explore different decimation values (guesses) along the main path and at very early steps, see Fig. \ref{fig:decision_tree}. These paths greatly improve the BP convergence speed and can run in independent parallel threads. We call our modified version of BPGD a \ac{GDG} decoder. Note that, despite its name, \ac{GDG} is a \emph{deterministic} decoder since both options are explored, instead of a randomly chosen value for the chosen VN.
We carefully optimize the maximum number of BP iterations and keep the number of paths small for GDG to be useful in the real-time setting. 


The remainder of the paper is organized as follows. BP and BP+OSD decoding methods are reviewed in Section \ref{sec:decoding}, followed by a review of the global decoding of circuit-level noise \cite{bravyi2023highthreshold} in Section \ref{sec:circuit_level_noise}. 
Details of sliding window decoding are given in Section \ref{sec:sliding_window} and of the GDG decoder in Section \ref{sec:GDG}. 
Additionally, in Appendix \ref{sec:data_qubit_noise} and \ref{sec:single_shot}, we apply both BP+OSD and GDG to data qubit noise and single-shot syndrome noise decoding to form comparisons. 

\section{Circuit-level noise decoding}
We follow the standard circuit-based depolarizing noise model \cite{high_threshold_fowler_2009}, which is also considered in \cite{bravyi2023highthreshold}. Each operation in the (repeated) syndrome measurement circuit, including CNOT gates, qubit initializations, measurements, and idle qubits, is subject to noise. 
One can imagine the gates occurring at integer timesteps starting at $1$; then the possible fault locations are half-integer timesteps for all qubits (i.e. between all gates), including ancillas. The decoding on the much simpler data qubit error model is discussed in Appendix \ref{sec:data_qubit_noise}, where it is assumed that the syndrome extraction circuit is noiseless and faults only occur at timestep $0.5$.

The circuit-level noise decoder outputs a Pauli correction operator, to be applied to the output data qubits, upon observing all the noisy syndromes from multiple rounds and taking into account all possible faults in the SM circuit. The decoding fails if the correction operator and the actual error differ by a non-trivial logical operator of the code.
Here we consider decoding using syndromes from X-type or Z-type measurements separately. This is enabled by the CSS structure \cite{CSS-CS,CSS-Steane} of \ac{BB} codes.

\subsection{Decoding}
\label{sec:decoding}
Consider the Tanner graph \cite{tanner_graph} associated with a binary linear code specified by a parity-check matrix (PCM) $\bH$. The Tanner graph is bipartite and consists of check nodes (CNs) and variable nodes (VNs) that each represent a row or a column of $\bH$, respectively. The presence of one at column $i$ and row $j$ of $\bH$ indicates an edge from VN $v_i$ to CN $c_j$. 
Associated to this edge are messages in both directions, $\mu_{i\to j}$ and $\mu_{j\to i}$. 
In the following, we use the concept of VN and column interchangeably. 
A binary variable can be associated to each VN, indicating a fault on the corresponding qubit or locations in the data qubit noise and circuit-level noise scenarios, respectively. 

In syndrome BP decoding, each CN $c_j$ receives a syndrome (a check value) $s_j\in\{0,1\}$, and we denote the entire vectors of syndromes $\bs$. 
Each VN $v_i$ receives as input a prior probability $p_i$ of it being flipped, with associated \acf{LLR} $\Lambda_i=\log\frac{1-p_i}{p_i}$.
We assume errors on VNs to be independent. 
For BP initialization (timestep $t=0$), the VN $v_i$ to CN $c_j$ messages is $\mu_{i\to j}^{(0)}=\Lambda_i$.

The message-passing algorithm proceeds iteratively, with one timestep consisting of first CN updates and then VN updates. 
At timestep $t$, the min-sum CN update rule computes, for each CN $c_j$, a message $\mu_{j\to i}$ to each of its neighboring VNs $v_i$, with $i\in \mathcal{N}(j)$, where 
\begin{equation}
\label{eq:min_sum}
    \mu_{j\to i}^{(t)}=(-1)^{s_j}\cdot\min\limits_{i'\in \mathcal{N}(j)\backslash i} |\mu_{i'\to j}^{(t-1)}| \prod\limits_{i'\in \mathcal{N}(j)\backslash i} \text{sign}(\mu_{i'\to j}^{(t-1)}).
\end{equation}
The VN update first calculates a posterior LLR $\Lambda_i^{(t)}$ for VN $v_i$ at timestep $t$ by summing up the original LLR $\Lambda_i$ with all incoming messages from its CN neighbors $\mathcal{M}(i)$, i.e.,
\begin{equation}
\label{eq:posterior}
    \Lambda^{(t)}_i=\Lambda_i + \sum\limits_{j'\in\mathcal{M}(i)}\mu_{i\leftarrow j'}^{(t)}.
\end{equation}
Then the message $\mu_{i\to j}^{(t)}$ is updated by subtracting the intrinsic message from the posterior, i.e.,
\begin{equation}
\label{eq:VN_update}
    \mu_{i\to j}^{(t)}=\Lambda_i^{(t)}-\mu_{i\leftarrow j}^{(t)}.
\end{equation}

Based on the posterior LLR at any timestep $t$, a decision for the estimated error $\hbe$ can be made \emph{locally} for each VN $i$
\begin{equation}
\label{eq:decision} 
\hat{e}_i=\begin{cases}
    0 & \text{if $\Lambda_i^{(t)} > 0$}\\
    1 & \text{if $\Lambda_i^{(t)} \leq 0$}.
    \end{cases}
\end{equation}
When the Tanner graph contains loops, the posterior LLR is only an approximated version of the true value; thus, it is possible that the BP's decision does not have the correct syndrome even after an infinite number of decoding iterations. 
Therefore, we stop BP as soon as $\hbe$ satisfies the syndrome equation $\bH\hbe=\bs$, or after some fixed number of iterations.

Since the prior LLRs on VNs need not be identical, which will be especially relevant in circuit-level decoding, we use the handy notation of the \emph{path metric} (PM). 
It is defined to be the sum of prior LLRs from the VNs that are estimated to one, if the estimation has the correct syndrome, or $\infty$ otherwise.
\begin{equation}
\label{eq:PM}
  \PM(\hbe) =
    \begin{cases}
\sum\limits_{i:\ \hat{e}_{i}=1}\Lambda_i & \text{if $\bH\hbe=\bs$,}\\
      \quad\quad\infty & \text{if $\bH\hbe\neq\bs$.}
    \end{cases}  
\end{equation}
A smaller path metric, therefore, corresponds to a higher probability of the estimated error pattern. 

In the later section, we will use a technique called VN decimation, which is to fix the value of a VN and remove it from the message-passing network. For example, if we choose to decimate $v_i$ to $0$, then this VN is no longer active and will be excluded from the updates of its CN neighbors. If we choose to decimate $v_i$ to one, the syndrome $s_j$ on all its CN neighbors $j\in\mathcal{M}(i)$ needs to be flipped, and the updated syndromes are used in Eq. (\ref{eq:min_sum}) for later CN updates.
In our implementation, instead of deleting a VN, we maintain a mask for the VN status. The messages from and to a decimated VN remain in the network, and all its CN neighbors $\mathcal{M}(i)$ ignore these \emph{stale} messages in their updates. 

In data qubit noise decoding of quantum LDPC codes, BP often outputs an error estimate $\hbe$ which does not satisfy the syndrome equation, this is usually termed non-convergence\footnote{Convergence and syndrome consistency are two different notions. The former does not imply the latter. However, we force the latter to imply the former by early stopping of BP once the syndrome equation is satisfied, in the sense that the estimation based on posterior LLR no longer changes.} and exhibits an error floor in logical error rates. One reason is due to the short loops present in the Tanner graph. The short to medium block-length QLDPC codes usually have a girth (the length of the shortest cycle) of four or six\footnote{Here we consider the PCMs $\bH_X$ and $\bH_Z$ for $X$ and $Z$ errors separately. See Table (1) of \cite{panteleev2021degenerate} for the girth of some QLDPC codes. For quaternary BP decoding on $(\bH_X,\bH_Z)$, four-cycles are unavoidable \cite{15years}.}. 

A few classical tricks for loopy BP can be employed to ameliorate this problem. 
A normalization/scaling factor $\alpha\leq 1$ can be multiplied to the right-hand side of Eq. (\ref{eq:min_sum}) to prevent over-amplified messages due to short cycles. The scaling factor is set to 1.0 in this work unless specified otherwise. Another trick is to change the scheduling method. Instead of all CNs applying the update rule in parallel in one BP iteration (flooding scheduling), only one CN (serial scheduling) or a fraction of CNs (layered scheduling) is updated before the next VN update. Serial scheduling usually reduces syndrome inconsistency with the same number of CN updates; however, those updates are sequential, which affects latency. In this work, we always use flooding scheduling within window decoding due to the potential latency constraint. 
Additionally, the (normalized) min-sum rule is used for the CN update, which is more hardware-efficient despite being an approximation. 

Apart from short loops, symmetric trapping sets \cite{perturbation,trapping_set_QLDPC} also complicate the decoding of QLDPC codes. 
Various post-processing methods have been proposed to break this symmetry, the most prominent being ordered-statistics decoding (OSD) \cite{panteleev2021degenerate,roffe2020osd-cs}. 
If BP fails to converge after some fixed number of iterations, the columns of $\bH$ are reordered according to the posterior LLRs of the VNs from low to high (the most-to-least likely of being flipped)\footnote{The binary BP+OSD proposed in Section 3.1 of \cite{panteleev2021degenerate} ranks columns according to reliabilities, i.e., the absolute value of posterior LLR, then solve the equation on least reliable columns by first fixing the rest to their BP hard decisions. \cite{roffe2020osd-cs} ranks according to likeliness of being flipped, i.e., the value of posterior LLR, which is the approach we follow here.}. 
Then the first rank($\bH$) \emph{linearly independent} columns are selected and used to find an error $\hbe$ satisfying the syndrome constraint $\bH\hbe= \bs$. 
The unselected values of $\hbe$, associated with the remaining columns of $\bH$, are set to zero. This is the zeroth-order OSD (denoted as BP+OSD-0).
The combination sweep heuristic \cite{roffe2020osd-cs} for order-$\lambda$ OSD (denoted as BP+OSD-CS$\lambda$) additionally searches over all weight-one configurations of all the unselected VNs, and weight-two patterns of the first $\lambda$ unselected VNs. BP+OSD may still exhibit an error floor, and the behavior is affected by the number of iterations, the scaling factor, and the scheduling used for the BP preprocessing \cite{stabilizer_inactivation}.

\subsection{Circuit-level noise}
\label{sec:circuit_level_noise}
In \cite{bravyi2023highthreshold}, along with the parity-check matrices for the \acf{BB} codes, a carefully designed gate ordering for the noisy SM circuit is additionally proposed. The SM is repeated for several rounds due to the unreliability of the measurements. 
A corresponding PCM can be constructed, see Fig. 14 of \cite{higgott2023sparse} for a graphical introduction to how such ``circuit codes'' are created.

Contrary to data qubit noise, where a VN denotes an error on a data qubit and a CN denotes a noiseless syndrome measurement result, for circuit-level noise a VN denotes a \emph{single fault} in the entire (multi-round) SM circuit while a CN denotes a detector, which is the XOR of the measurement results of a given parity-check from two consecutive rounds. 
The presence of one in column $i$ and row $j$ of the circuit code PCM means the $i^{th}$ fault mechanism triggers the $j^{th}$ detector. 

As an example, when repeating the SM circuit for $R$ rounds, consider one syndrome check $c$ whose measurement results from oldest to most recent in time are $m_1,m_2,\dots,m_R$. The resulting detectors associated with this check $c$ are checking $m_1, m_2\oplus m_1, m_3\oplus m_2,\dots, m_{R}\oplus m_{R-1}$, which will be zero in the absence of any faults. 
Now consider the VN that denotes a single measurement flip in round $r>2$ on this particular check $c$. 
This VN has degree two because it triggers both detectors $m_{k-1}\oplus m_k$ and $m_{k+1}\oplus m_{k}$ to one, and it does not trigger detectors associated to other checks. 
In Fig. \ref{fig:visualization} showing the circuit code PCM $\bH_{circ}$, we group rows (detectors) by round. The top block of rows (the rows that the top $\bH_0$ occupies) are detectors like $m_1$ associated to all checks, then the next block of rows are $m_2\oplus m_1$ like detectors, and so on. 
The bottom block of rows are the most recent detector values. 

Further in Fig. \ref{fig:visualization}, one can see that there are two kinds of faults.
One kind triggers detectors in the current round (block of rows) only; these faults form the columns of $\bH_0$. 
An example is a fault on a data qubit immediately prior to a round of syndrome extraction, which causes check values to change only in that round. 
The other kind of fault triggers detectors in consecutive two rounds, constituting the columns of 
$\begin{bsmallmatrix}
    \bH_1 \\ \bH_2
\end{bsmallmatrix}$.
Single measurement faults are an example, as well as the various faults in the middle of the SM circuit. The latter are captured by only a subset of the syndrome checks in the occurring round, but can be fully captured by all subsequent rounds. Using the same argument, one can see that no single fault triggers detectors from more than two rounds, thanks to the sparsification created by the XOR operation.


Fault mechanisms can be combined if they trigger the same set of detectors, have the same noiseless syndrome at the end, and induce the same logical error. A simple example is an X-flip on the target qubit before and after a CNOT gate. 
The columns of such equivalent fault mechanisms are merged, and a new prior probability is calculated for the new VN. In this linearized model \cite{bravyi2023highthreshold}, the possible correlation between columns is ignored, meaning the different single fault mechanisms are assumed to happen independently. Given all the detector values and the final noiseless syndrome, the decoder aims to find the most probable set of faults that explains the measurement result. Then the correction operator on the data qubits can be determined by adding up the  effects on the output data qubits caused by each selected fault.

Several undesired structures in the circuit code Tanner graph harm its BP performance. An X-flip on the control qubit propagates to both control and target qubits after the CNOT. Consider the triplet of columns representing these three single-fault mechanisms. The XOR of any two columns equals the other, which creates numerous short cycles.
The aforementioned measurement fault triggers two detectors and corresponds to a VN of degree two, which is known to cause high error floors in BP decoding \cite{window_SC}. 
To alleviate the syndrome inconsistency problem, BP+OSD is employed in \cite{bravyi2023highthreshold} to perform global decoding of the fault mechanisms that happened in $d$ rounds of noisy syndrome measurements, where $d$ is the code distance. This method is not likely to fulfill the latency constraint, due to the large PCM of the circuit code and the high worst-case runtime of OSD.

In this work, we use the sliding window decoder that is suitable for handling measurement data coming in a streaming fashion.
In fact, the XOR of check results from consecutive two SM rounds can be seen as naturally creating a spatially-coupled LDPC (SC-LDPC)-like code, and the sliding window decoder is a standard decoding technique for \emph{classical} SC-LDPC codes \cite{window_SC}.

\section{Sliding Window Decoding}
\label{sec:sliding_window}
We demonstrate the sliding-window decoding of the bivariate bicycle (BB) code family implemented using the proposed \acf{SM} circuit in Fig. 7 of \cite{bravyi2023highthreshold}. X-type and Z-type check operators are decoded separately as in \cite{bravyi2023highthreshold}, though circuit-level depolarizing noise is assumed. 
Correlations between X and Z errors could be considered; however, we find that in this case fewer fault mechanisms can be combined, and the PCM of the circuit code becomes significantly larger. In particular, we find the resulting PCM has roughly eight times more columns.

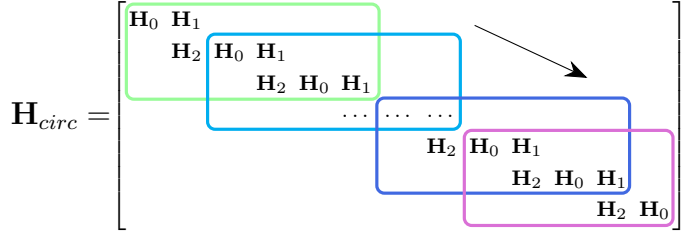
\begin{figure}
    \centering
    \usetikzlibrary{matrix}
\usetikzlibrary{positioning}
\usetikzlibrary {arrows.meta}
\definecolor{orchid}{HTML}{DA70D6}
\definecolor{palegreen}{HTML}{98FB98}
\definecolor{royalblue}{HTML}{4169E1}

\tikzstyle{box} = [draw,rounded corners=.1cm,inner sep=5pt,minimum height=4.5em, text width=10.7em, align=center,very thick] 

\tikzstyle{box3} = [draw,rounded corners=.1cm,text width=9.2em,minimum height=4.5em,align=center,very thick]

\begin{tikzpicture}[
    every left delimiter/.style={xshift=.65em},
    every right delimiter/.style={xshift=-.65em},
  ]
\footnotesize
  \matrix (m) [matrix of nodes,left delimiter={[},right delimiter={]}, row sep=0\pgflinewidth, column sep=0.2\pgflinewidth,
nodes={rectangle, 
               minimum height=1.5em, minimum width=2em,
                      anchor=center, 
                      inner sep=0pt, outer sep=0pt},
                      ]
  {
    $\bH_0$ & $\bH_1$ \\
    & $\bH_2$ & $\bH_0$ & $\bH_1$ \\
    & & & $\bH_2$ & $\bH_0$ & $\bH_1$  \\
    & & & & & $\dots$ & $\dots$ & $\dots$  \\
    & & & & & & & $\bH_2$ & $\bH_0$ & $\bH_1$  \\
    & & & & & & & & & $\bH_2$ & $\bH_0$ & $\bH_1$  \\
    & & & & & & & & & & & $\bH_2$ & $\bH_0$  \\
  } ;

\node [color=palegreen,box](w1)  at (-1.95,0.85){}; 
 
\node [color=cyan,box](w2)  at (-0.87,0.45){};

\node [color=royalblue,box](w3)  at (1.38,-0.4){};
 
\node [color=orchid,box3](w4)  at (2.25,-0.83){};


\node (h) at (-4.5,0) {\large $\bH_{circ}=$};

\draw [-{Stealth[length=3mm]}] (1, 1.2) -- (2.5, 0.5);

\end{tikzpicture}
    \vspace*{-4mm}
    \caption{(3,1) sliding window decoding of the circuit code PCM $\bH_{circ}$. For the BB code family with circuit in \cite{bravyi2023highthreshold}, $\bH_0$ has shape $w\times 3w$, $\bH_1, \bH_2$ both have shape $w\times 9w$, where $w=N/2$ is the number of detectors used in one round for a block-length $N$ code.}
    \label{fig:visualization}
\end{figure}

We implement a memory experiment in the Z basis in Stim \cite{stim}, using the repeated SM circuit from Fig. 7 of \cite{bravyi2023highthreshold}. 
We find that this family of codes possesses a general description of their corresponding circuit codes in terms of the shape of $\bH_0,\bH_1,\bH_2$, see the caption of Fig~\ref{fig:visualization}. 

In the (3,1)-sliding window decoding of this circuit code, three rounds of detector values are handled in a window and the window moves downward by one round each time. 
Inside each window, an inner decoder is applied, e.g., BP+OSD-CS10 or GDG. The decisions for the columns not contained in the next window are committed, and the detector values for the next window are updated. 

For example, assume detector values $\bs_1,\bs_2,\bs_3,\dots$ are observed, where each $\bs_i$ is associated to the $i^{th}$ block of rows. We first use the first window (green) PCM $\bH_{win}$ to solve the following syndrome equation (and all subsequent ones) over the binary field,
\begin{equation}
\label{eq:win_eq}
    \begin{bmatrix}
    \bH_0 & \bH_1 &  &  &  &\\ 
          & \bH_2 & \bH_0 & \bH_1 & &\\
          &       &       & \bH_2 & \bH_0 & \bH_1
    \end{bmatrix}
    \begin{bmatrix}
        \hbe_0\\
        \hbe_1\\
        \vdots
    \end{bmatrix}=
    \begin{bmatrix}
        \bs_1\\
        \bs_2\\
        \bs_3
    \end{bmatrix}.
\end{equation}
Here $\hbe_0$ and $\hbe_1$ are the estimated VN patterns associated with the columns of leftmost $\bH_0$ and $\bH_1$ respectively, and they are the columns that we commit to before moving to the next window. 

It is clear that if an inner decoder gives a solution that satisfies this equation, then 
\begin{equation}
\label{eq:win_partial}
    \begin{bmatrix}\bH_0 & \bH_1\end{bmatrix}\begin{bmatrix}\hbe_0 \\ \hbe_1\end{bmatrix} = [\bs_1]
\end{equation}
is naturally satisfied. \ac{BP+OSD} can always find such a solution, however, by applying plain BP alone to this window decoding problem, we empirically observed that the convergence is weak for the \ac{BB} code family. Importantly, if Eq. (\ref{eq:win_partial}) is not satisfied, then the overall syndrome equation
\begin{equation}
\label{eq:overall}
    \bH_{circ}   \begin{bmatrix}\hbe_0\\\hbe_1\\ \smash{\vdots}\end{bmatrix}=\begin{bmatrix}\bs_1\\\bs_2\\ \smash{\vdots}\end{bmatrix}
\end{equation}
can never be satisfied, regardless of subsequent window decoding. 

By committing to $\hbe_0$ and $\hbe_1$, we need to update $\bs_2$ used for the next window by $\bs_2'=\bs_2+\bH_2\hbe_1$, then use $\bs'_2,\bs_3,\bs_4$ to decode the next window. 
If all partial equations like Eq. (\ref{eq:win_partial}) hold with respect to newly committed columns and (updated) syndrome, then Eq. (\ref{eq:overall}) will also be satisfied, and vice versa.
When employing our GDG decoder for window decoding, we always count Eq. (\ref{eq:overall}) failures into logical errors, besides requiring the overall logical errors caused by the chosen VNs to match the one that Stim \cite{stim} returns. 

More concretely, consider a simulation for protection of the logical $Z$ information in a $\llbracket N,K,d\rrbracket$ code that encodes $K$ logical qubits. 
A logical $Z$ observable matrix \cite{higgott2023sparse} $\mathbf{L}$ consisting of $K$ rows and the same number of columns as $\bH_{circ}$ can be constructed as follows. Imagine adding $K$ checks associated to the logical operators to the very end of the multi-round SM circuit. For the single-fault represented by column $i$ of $\bH_{circ}$, propagate it to the end of the SM circuit, just before the logical checks. Put a one in the $i^{th}$ columns and $j^{th}$ row of $\mathbf{L}$ if this final error string does not commute with (triggers) the logical check $j$. For numerical simulation, Stim \cite{stim} provides $K$ bits $l_1,\dots,l_K$ indicating the values of the final logical checks, along with the detector values mentioned earlier. 
We decode using the detector values only and determine if our final estimation satisfies
\begin{equation}
\label{eq:logical}
    \mathbf{L}\begin{bmatrix}\hbe_0\\\hbe_1\\ \smash{\vdots}\end{bmatrix}=\begin{bmatrix}l_1\\\smash{\vdots}\\l_K\end{bmatrix}
\end{equation}
and Eq. (\ref{eq:overall}). The decoding is deemed successful if both equations are satisfied.

In classical SC-LDPC code decoding, BP is used on each window, and usually a small number of iterations of BP already leads to convergence. As mentioned above, this is not the case for $\bH_{circ}$ of the BB family. Therefore, we apply BP+OSD-CS10 to each window and benchmark the performance against the global decoding over distance $d$ rounds of SM, see Fig.~\ref{fig:LER}. 

Though each window deals with a much smaller PCM compared to global decoding over $d$ rounds, fulfilling the stringent latency constraint is still challenging. 
Before proceeding to the next section where we introduce the GDG decoder, we make a few comments on the windowed approach.

Each window is extremely wide, i.e., the number of columns is much larger than the number of rows. To improve efficiency, we merge the columns of the bottom right (full-rank) matrix $\bH_1$ in each window to an identity matrix and recalculate priors. Secondly, inspired by OSD-0, we notice that certain columns can be dropped (decided to zero) after BP posterior LLR ranking. 
For a window of $w'$ rows, instead of testing for linear dependency and choosing the first $\text{rank}(\bH)\approx w'$ columns as in OSD-0, we \emph{skip the test} and simply choose the first $2w'$ columns for post-processing. This method is always used together with GDG in the next section. 
As it turns out, the first $2w'$ columns are almost always enough for the syndrome to be in their span. This means OSD-0 will decimate the remaining columns to zero as well.

It is also observed that running BP+OSD (not just OSD) on the remaining $w'\times 2w'$ PCM can slightly improve the performance of decoding on the original PCM. The improvement is not shown in this paper for brevity but is available online. Restricting OSD-CS10 to the first $2w'$ columns after ranking improves worst-case runtime as well, since an order-$\lambda$ combination-sweep OSD tests weight-one patterns for \emph{all} columns, not just the first $\lambda$ unchosen columns.

\section{Guided Decimation Guessing Decoding}
\label{sec:GDG}
Decimation \cite{mezard, montanari, SC_BP_decimation} is a technique to improve convergence of iterative decoding by sequentially fixing VNs via hard decisions. It was recently used by \cite{yao2023BPGD} for decoding QLDPC codes subject to data qubit noise. 
There, after some BP iterations, the VN with the largest \emph{absolute value} of posterior LLR (the most reliable) is decimated according to its sign. This process is repeated for $R$ steps. However, if $R$ is limited to $50$ for the blocklength $882$ code, an error floor is observed at low physical error rates, see Fig.~4 of \cite{yao2023BPGD}. 
When investigating the cause, we found that in most of the steps the selected VN is decimated to zero, because the most reliable posterior LLR is usually positive, due to most messages being positive at low error rates.

This behavior inspired us to instead select the VN with the smallest posterior LLR (the most likely to flip) and again decimate according to the sign of the LLR. In this way, the decoder becomes more effective in the low error rate regime.
The rationale is as follows. Imagine an error $\be$ with Hamming weight $\wt(\be)$ occurred. 
Then, we expect the smallest posterior LLR at each step to be negative and a one to be decided for the decimated VN. 
After $\wt(\be)$ number of decimation steps, BP should have converged and recovered  $\be$. 
The average $\wt(\be)$ decreases when the physical error rate gets lower, and BP thus should take fewer steps to converge.

In numerical experiments, however, we find that the smallest posterior LLR is not necessarily negative at each step. 
Convergence can nevertheless usually be achieved by always following the sign and allowing decimation to proceed until possibly no VNs left, but the downside is that the estimated error sometimes has a very large weight. 
This effect accumulates when sliding the window multiple times and manifests in the high logical error rate in the end. 
On the other hand, the runtime is not guaranteed when taking a lot of steps to find a large-weight solution. To address the challenge of finding a low-weight solution that has the correct syndrome in a fixed number of steps, we propose two key techniques to increase convergence speed. One is to select and decimate VN based on their posterior LLR history so that the guidance we follow is more reliable. The other is to employ \emph{guessing}, i.e., trying both decimation values.

\subsection{History-based decision}
When plotting the LLR posterior against the number of BP iterations, we find that some VNs exhibit an oscillating behavior. They oscillate at a period of four, which is the girth of the circuit code Tanner graph. Therefore, we use a buffer to store the posterior LLRs from the latest four iterations, and always use this history to make decimation decisions.

We use a ternary mask $vn\_status$ for each VN with $-1$ indicating this VN is not yet decimated (active), and $0/1$ for being decimated to $0/1$ (inactive). 
After decimating a VN to 1, the syndromes of its neighboring CNs are flipped. 
When running BP, the inactive VNs no longer update messages, and CNs simply ignore stale messages left behind by their inactive VN neighbors. 
For the time being, assuming we are in low error mode in Alg. \ref{alg:select_vn}, where all the active VNs with degrees larger than two enter the VN selection. This mode is used for physical error rate $p\leq 0.002$ for all codes in Fig. \ref{fig:LER}. For larger $p$, we find it beneficial to leave the low error mode and immediately hard-decide VNs with a very positive or negative history to zero or one, without including them into the VN selection.

\begin{algorithm}
\hspace*{\algorithmicindent} \textbf{Input} Depth $D$\\
\hspace*{\algorithmicindent} \textbf{Output} Chosen VN index $g$, favored guess value $f$
\caption{Select a VN to decimate and compute the favored value based on LLR history}
\label{alg:select_vn}
\begin{algorithmic}[1]
\Procedure{select-VN}{$D$}
\State $L_{min}\gets \infty$
\State $g_{min}\gets -1$
\State $\tilde{L}_{min}\gets \infty$\Comment{for all negative LLR history}
\State $\tilde{g}_{min}\gets -1$
\ForAll{VN $v_i$}
\If {$vn\_status[i]\neq -1$}\Comment{decimated}
\State {\textbf{continue}}
\EndIf
\If {deg$(v_i)\leq 2$}
\State {\textbf{continue}}
\EndIf

\State {// history: last 4 iter. of Eq. (\ref{eq:posterior})}
\State {$\mathbf{\Lambda}_{hist}=[\Lambda_{i}^{(t-3)},\Lambda_{i}^{(t-2)},\Lambda_{i}^{(t-1)},\Lambda_{i}^{(t)}]$}
\State $L_{sum}\gets$ sum$(\mathbf{\Lambda}_{hist})$
\If {not in low error mode}
\State {$b\gets$ \textsc{AGG-DEC}($\mathbf{\Lambda}_{hist}$, $D$)}\Comment{Alg. \ref{alg:aggressive_decimation}}
\If {$b\neq -1$} \Comment{$\mathbf{\Lambda}_{hist}$ reliable}
\State $vn\_status[i]\gets\hbe[i]\gets b$
\State {\textbf{continue}} \Comment{$v_i$ quits selection}
\EndIf
\EndIf
\If {$L_{sum} < L_{min}$}
\State {$L_{min}\gets L_{sum}$}
\State {$g_{min} \gets i$}
\EndIf
\If {$\mathbf{\Lambda}_{hist}\leq 0$ \textbf{and} $L_{sum}<\tilde{L}_{min}$}
\State {$\tilde{L}_{min} \gets L_{sum}$}
\State {$\tilde{g}_{min} \gets i$}
\EndIf
\EndFor

\State {// select one VN and favored value for it to guess}

\If{$\tilde{g}_{min}\neq -1$}
\State {$g\gets \tilde{g}_{min}$}
\State {$f\gets 1$}
\Else 
\State {$g\gets g_{min}$}
\State {$f\gets \text{sign}(L_{min})$}
\EndIf
\State \Return{$g$, $f$}
\EndProcedure
\end{algorithmic}
\end{algorithm}

In Alg. \ref{alg:select_vn}, if there are any VNs whose four most recent LLRs are all negative, then the VN with the most negative sum is chosen and decimated to one. In the absence of such VNs, we decimate the VN with the smallest sum of its four most recent LLRs to the value indicated by the sign of the sum. 
In particular, we also skip VNs whose degree is less than three, as this was found to improve performance. 
We believe the reason is that degree two or one VNs are just relay or observer nodes \cite{cammerer2018sparse} and themselves do not carry reliable information. 
In the circuit code window PCM $\bH_{win}$, the degree one and two VNs stand for the measurement faults, which should not be decoded to one with priority.

\subsection{Guessing}
We call the above decision path the \emph{main branch} and it is shown in red in Fig. \ref{fig:decision_tree}. 
When limiting the main branch to $25$ steps and employing it as the inner window decoder for the $N=144$ code, we observe error suppression, i.e., the logical error rate per round can be smaller than the physical error rate. 

We can further improve the performance by \emph{guessing}. 
That is, for a small number of variable nodes, we consider both possible decimated values and perform ensemble decoding.
When closely investigating the LLR history, we observed that sometimes the history of the selected VN oscillates around zero. 
This naturally leads us to explore the \emph{side branches} (blue squiggles in \Cref{fig:decision_tree}), which are decision paths that directly split off from the main branch. 
These decimate a VN contrary to the sign of the LLR history only at \emph{one point}, where the branch splits, and all subsequent decisions follow the sign. 
These side branches improve convergence in ensemble decoding significantly, and by incorporating them into the inner window decoder, we achieve BP+OSD-0 performance. Finally, inspired by the pattern exploration of unchosen VNs in high-order OSD, we add the \emph{tree branches} (green squiggles in \Cref{fig:decision_tree}) emanating from the leaves of a depth-4 \emph{guessing tree}\footnote{Following line 22-24 of Alg. \ref{alg:BPGDG}, the id labeled for tree branches in Fig. \ref{fig:decision_tree} is only a demonstration for a guessing tree where the favored value at each step is one, which is not always the case in reality.}. They explore all the decimation values of the first four decisions. After splitting, they all proceed following the sign. The tree branches help in finding a more probable (smaller path metric) estimated error pattern and they bridge us to BP+OSD-CS10 performance. We name the resulting decoder a guided decimation guessing (GDG) decoder.

\begin{figure}
    \centering
    \begin{tikzpicture}[every node/.style={circle, draw, fill, inner sep=0pt}, 
          level distance=0.5cm, sibling distance=5cm,
          level 1/.style={sibling distance=4cm},
          level 2/.style={sibling distance=2cm},
          level 3/.style={sibling distance=1cm},
          edge from parent path={(\tikzparentnode) -- (\tikzchildnode)}]]
    \node[red, label={[text=red]above right:1}] (1) {1}
        child {node (2) {2}
            child {node (4) {4}
                child {node (8) {8}}
                child {node (9) {9}}
            }
            child {node (5) {5}
                child {node (10) {0}}
                child {node (11) {0}}
            }
        }
        child {node[red, label={[text=red]above right:2}] (3) {3}
            child {node (6) {6}
                child {node (12) {0}}
                child {node (13) {0}}
            }
            child {node[red, label={[text=red]right:3}] (7) {7}
                child {node (14) {0}}
                child {node[red, label={[text=red]right:4}] (15) {0} 
                    child [right] {node[red, label={[text=red]right:5}] (16) {0}
                    child [right] {node[red, label={[text=red]right:6}] (17) {0}
                    child [left] {node[red, label={[text=red]left:7}] (18) {0}
                    child [right] {node[red, label={[text=red]right:8}] (19) {0}
                    child [left] {node[red, label={[text=red]left:9}] (20) {0}
                    child [right] {node[red, label={[text=red]right:10}] (21) {0}
                    }}}}}}
                }
            }
        };
        \draw[red, line width=2pt] (1) -- (3);
        \draw[red, line width=2pt] (3) -- (7);
        \draw[red, line width=2pt] (7) -- (15);
        \draw[red, line width=2pt] (15) -- (16);
        \draw[red, line width=2pt] (16) -- (17);
        \draw[red, line width=2pt] (17) -- (18);
        \draw[red, line width=2pt] (18) -- (19);
        \draw[red, line width=2pt] (19) -- (20);
        \draw[red, line width=2pt] (20) -- (21);
        \draw[red, line width=2pt] (1.north) -- (0,0.5);
        \draw[green, line width=1pt,decorate,decoration={snake,amplitude=.4mm,segment length=2mm}] (8) -- +(0.4,-0.8);
        \draw[green, line width=1pt,decorate,decoration={snake,amplitude=.4mm,segment length=2mm}] (8) -- +(-0.4,-0.8);

        \node[green, below right=8mm and -11mm of 8.south, rectangle, draw=none, fill=none] {id\ \ $15$\ \ $14$ $13$\ \ $12$ $11$\ \ $10$ $9$\ \ \ $8$\ \ $7$\quad\ $6$\ \ $5$\quad$4$\ \ $3$\quad\ $2$};

        \draw [decorate,decoration={brace,amplitude=5pt,mirror,raise=4ex},green] ([yshift=-5mm,xshift=-4mm]8.south) -- ([yshift=-5mm,xshift=4mm]14.south) node[rectangle,midway,green,yshift=-12mm,inner sep=0pt, fill=none, draw=none]{$D_{max}=4+10$};

        \draw[green, line width=1pt,decorate,decoration={snake,amplitude=.4mm,segment length=2mm}] (9) -- +(0.4,-0.8);
        \draw[green, line width=1pt,decorate,decoration={snake,amplitude=.4mm,segment length=2mm}] (9) -- +(-0.4,-0.8);
        \draw[green, line width=1pt,decorate,decoration={snake,amplitude=.4mm,segment length=2mm}] (10) -- +(0.4,-0.8);
        \draw[green, line width=1pt,decorate,decoration={snake,amplitude=.4mm,segment length=2mm}] (10) -- +(-0.4,-0.8);
        \draw[green, line width=1pt,decorate,decoration={snake,amplitude=.4mm,segment length=2mm}] (11) -- +(0.4,-0.8);
        \draw[green, line width=1pt,decorate,decoration={snake,amplitude=.4mm,segment length=2mm}] (11) -- +(-0.4,-0.8);
        \draw[green, line width=1pt,decorate,decoration={snake,amplitude=.4mm,segment length=2mm}] (12) -- +(0.4,-0.8);
        \draw[green, line width=1pt,decorate,decoration={snake,amplitude=.4mm,segment length=2mm}] (12) -- +(-0.4,-0.8);
        \draw[green, line width=1pt,decorate,decoration={snake,amplitude=.4mm,segment length=2mm}] (13) -- +(0.4,-0.8);
        \draw[green, line width=1pt,decorate,decoration={snake,amplitude=.4mm,segment length=2mm}] (13) -- +(-0.4,-0.8);
        \draw[green, line width=1pt,decorate,decoration={snake,amplitude=.4mm,segment length=2mm}] (14) -- +(0.4,-0.8);
        \draw[green, line width=1pt,decorate,decoration={snake,amplitude=.4mm,segment length=2mm}] (14) -- +(-0.4,-0.8);
        \draw[blue, line width=1pt,decorate,decoration={snake,amplitude=.4mm,segment length=2mm}] (15) -- +(-0.4,-0.8);
        \draw[blue, line width=1pt,decorate,decoration={snake,amplitude=.4mm,segment length=2mm}] (16) -- +(-0.4,-0.8);
        \draw[blue, line width=1pt,decorate,decoration={snake,amplitude=.4mm,segment length=2mm}] (17) -- +(0.4,-0.8);
        \draw[blue, line width=1pt,decorate,decoration={snake,amplitude=.4mm,segment length=2mm}] (18) -- +(-0.4,-0.8);
        \draw[blue, line width=1pt,decorate,decoration={snake,amplitude=.4mm,segment length=2mm}] (19) -- +(0.4,-0.8);
        \draw[blue, line width=1pt,decorate,decoration={snake,amplitude=.4mm,segment length=2mm}] (20) -- +(-0.4,-0.8);
        \draw[blue, line width=1pt,decorate,decoration={snake,amplitude=.4mm,segment length=2mm}] (21) -- +(-0.4,-0.8);
        \node[rectangle,blue,below left=6mm and 5mm,inner sep=0pt, fill=none, draw=none] at (21.west) {$D_{max}=10+10$};
        \draw[red, line width=2pt,decorate,decoration={snake,amplitude=.4mm,segment length=2mm}] (21) -- +(0.4,-0.8) node[rectangle,draw=none, fill=none, below left=2mm and -3mm] {$D_{max}=25$};

\end{tikzpicture}
    \vspace*{-2mm}
    \caption{The decision tree for the BPGDG algorithm. VN selection and decimation are made after each step, where one step is defined to be six BP iterations. The red path is the main branch. The solid dots (red or black) are the only places where guessing is allowed. No splitting from the main branch into side branches (blue) after reaching depth 10. No guessing for the tree branches (green) after depth 4. The main branch terminates after 25 steps ($D_{max}=25$), regardless of convergence. The side branches are allowed to run 10 more steps after their split-off at depth $D_{splitt}$, i.e., $D_{max}=D_{splitt}+10$. The tree branches are also allowed to run 10 more steps after splitting from their neighbors at depth $4$, which means $D_{max}=4+10$ for all tree branches.}
    \label{fig:decision_tree}
\end{figure}
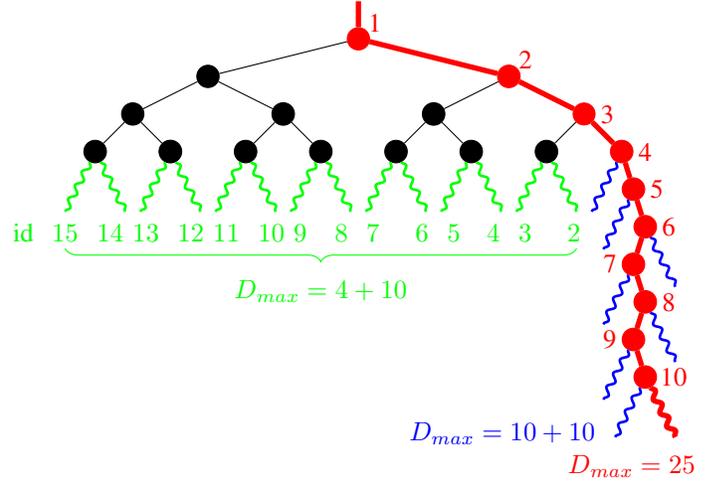
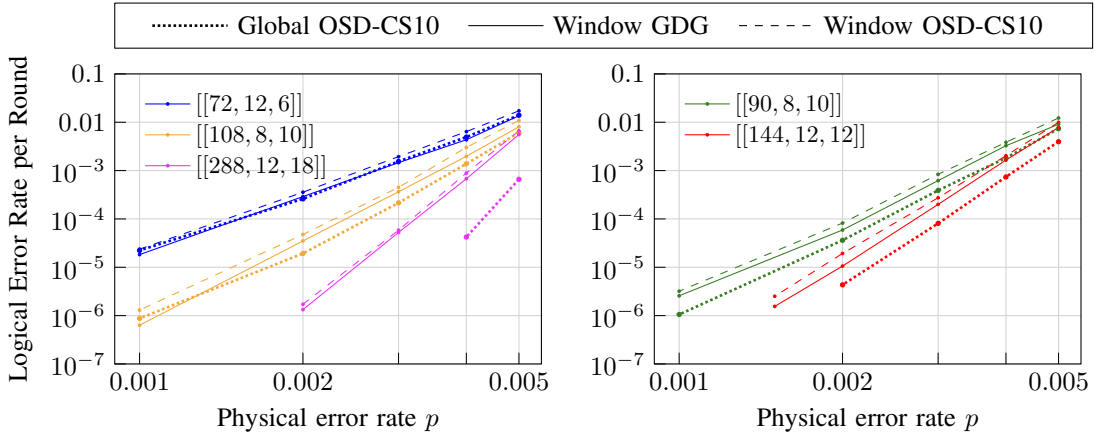
\begin{figure*}[htb]
\centering
\begin{tikzpicture}
\definecolor{blueviolet}{RGB}{138,43,226}
\definecolor{darkgray176}{RGB}{176,176,176}
\definecolor{lightgray211}{RGB}{211,211,211}
\definecolor{darkorange25512714}{RGB}{255,127,14}
\definecolor{dodgerblue}{RGB}{30,144,255}
\definecolor{gold}{RGB}{255,215,0}
\definecolor{red}{RGB}{255,0,0}
\definecolor{limegreen}{RGB}{50,205,50}
\definecolor{steelblue31119180}{RGB}{31,119,180}
\definecolor{teal}{RGB}{0,128,128}
\definecolor{orange}{RGB}{242,169,59}
\definecolor{olivegreen}{RGB}{55,126,34}
\definecolor{darkviolet}{RGB}{153,50,204}
\definecolor{neonpink}{RGB}{234,51,247}

\begin{groupplot}[
group style={
  group name=myplot,
  group size= 2 by 1, 
  horizontal sep=1.5cm
},
width=0.4\textwidth,
height=0.3\textwidth,
legend cell align={left},
legend columns=5,
transpose legend,
legend style={
  fill opacity=0,
  draw opacity=1,
  text opacity=1,
  at={(0.05,0.97)},
  anchor=north west,
  draw=none,
  fill=none,
  nodes={scale=0.9, transform shape}
},
log basis y={10},
tick align=inside,
tick pos=left,
x grid style={lightgray211},
xlabel={Physical error rate \(\displaystyle p\)},
xmajorgrids,
xmode=log,
xmin=0.0009, xmax=0.0055,
xtick={1e-3,2e-3,3e-3,4e-3,5e-3},
xticklabels={
  \(\displaystyle {0.001}\),
  \(\displaystyle {0.002}\),
  \(\displaystyle \),
  \(\displaystyle \),
  \(\displaystyle {0.005}\),
},
minor xtick={0.05,0.06,0.07,0.08,0.09},
xminorgrids=true,
scaled x ticks=false,
xtick style={color=black},
ylabel={Logical Error Rate},
y grid style={lightgray211},
ymajorgrids,
ymin=1e-7, ymax=0.1,
ymode=log,
ytick style={color=black},
ytick={1e-7,1e-6,1e-5,0.0001,0.001,0.01,0.1,1},
yticklabels={
  \(\displaystyle {10^{-7}}\),
  \(\displaystyle {10^{-6}}\),
  \(\displaystyle {10^{-5}}\),
  \(\displaystyle {10^{-4}}\),
  \(\displaystyle {10^{-3}}\),
  \(\displaystyle {0.01}\),
  \(\displaystyle {0.1}\),
  \(\displaystyle {1}\),
},
]

\nextgroupplot[ylabel={Logical Error Rate per Round},
               ]

\addplot [blue, dashed, mark=*, mark size=0.5, mark options={solid}, forget plot]
table [x=P, y=LER] {n72_k12_d6_W3_F1_osd.dat};
\addplot [blue, densely dotted, line width=1.0, mark=*, mark size=0.5, mark options={solid}, forget plot]
table [x=P, y=LER] {n72_k12_d6.dat};
\addplot [blue, mark=*, mark size=0.5, mark options={solid}]
table [x=P, y=LER] {n72_k12_d6_W3_F1_guessing.dat};
\addlegendentry{$[[72,12,6]]$}

\addplot [orange, densely dotted, line width=1.0, mark=*, mark size=0.5, mark options={solid}, forget plot]
table [x=P, y=LER] {n108_k8_d10.dat};
\addplot [orange, dashed, mark=*, mark size=0.5, mark options={solid}, forget plot]
table [x=P, y=LER] {n108_k8_d10_W3_F1_osd.dat};
\addplot [orange, mark=*, mark size=0.5, mark options={solid}]
table [x=P, y=LER] {n108_k8_d10_W3_F1_guessing.dat};
\addlegendentry{$[[108,8,10]]$}

\addplot [neonpink, densely dotted, line width=1.0, mark=*, mark size=0.5, mark options={solid}, forget plot]
table [x=P, y=LER] {n288_k12_d18.dat};
\addplot [neonpink, dashed, mark=*, mark size=0.5, mark options={solid}, forget plot]
table [x=P, y=LER] {n288_k12_d18_W3_F1_osd.dat};
\addplot [neonpink, mark=*, mark size=0.5, mark options={solid}]
table [x=P, y=LER] {n288_k12_d18_W3_F1_guessing.dat};
\addlegendentry{$[[288,12,18]]$}


\coordinate (left) at (rel axis cs:1,0);
\nextgroupplot[ylabel={}]

\addplot [olivegreen, densely dotted, line width=1.0, mark=*, mark size=0.5, mark options={solid}, forget plot]
table [x=P, y=LER] {n90_k8_d10.dat};
\addplot [olivegreen, dashed, mark=*, mark size=0.5, mark options={solid}, forget plot]
table [x=P, y=LER] {n90_k8_d10_W3_F1_osd.dat};
\addplot [olivegreen, mark=*, mark size=0.5, mark options={solid}]
table [x=P, y=LER] {n90_k8_d10_W3_F1_guessing.dat};
\addlegendentry{$[[90,8,10]]$}

\addplot [red, densely dotted, line width=1.0, mark=*, mark size=0.5, mark options={solid}, forget plot]
table [x=P, y=LER] {n144_k12_d12.dat};
\addplot [red, dashed, mark=*, mark size=0.5, mark options={solid}, forget plot]
table [x=P, y=LER] {n144_k12_d12_W3_F1_osd.dat};
\addplot [red, mark=*, mark size=0.5, mark options={solid}]
table [x=P, y=LER] {n144_k12_d12_W3_F1_guessing.dat};
\addlegendentry{$[[144,12,12]]$}

\coordinate (right) at (rel axis cs:1,0);
\end{groupplot}
\draw (0,4.8) -- (13,4.8) -- (13,4.2) -- (0,4.2) -- (0,4.8);
\draw [line width=1.0, densely dotted] (0.5,4.5) -- (1.5,4.5) node[right, pos=1.0] {Global OSD-CS10}; 
\draw (4.7,4.5) -- (5.7,4.5) node[right, pos=1.0] {Window GDG};
\draw [dashed] (8.3,4.5) -- (9.3,4.5) node [right, pos=1.0] {Window OSD-CS10};

\end{tikzpicture}
\vspace*{-4mm}
\caption{Logical error rate per round. The syndrome measurement is repeated $d$ rounds for a distance $d$ code. Dotted lines are global decoding over $d$ rounds using BP (1000 iterations) + OSD-CS10. Dashed lines are (3,1)-sliding window decoding where each window uses BP (200 iterations) + OSD-CS10 as the inner decoder. Solid lines use GDG as the inner decoder in (3,1)-sliding window. For GDG, low error mode (no aggressive decimation) is used for $p\leq 0.002$ for all codes.}
\label{fig:LER}
\end{figure*}
\begin{algorithm}
 \hspace*{\algorithmicindent} \textbf{Input} Parity-check matrix $\bH$, syndrome $\bs$\\
 \hspace*{\algorithmicindent} \textbf{Output} Estimated error $\hbe$
\caption{Belief Propagation Guided Decimation Guessing}\label{alg:BPGDG}
\begin{algorithmic}[1]
\ForAll {decision paths in Fig. \ref{fig:decision_tree}}
\ForAll {VN $v_i$} \Comment{initialization status mask}
\State {$vn\_status[i]\gets -1$} \Comment{not yet decimated}
\EndFor
\For{$D\gets 1,\dots,D_{max}$}
    \State {$\hbe\gets$ run BP for 6 iterations}
    \If{$\bH\hbe=\bs$}
        \State {calculate path metric for $\hbe$ with Eq. (\ref{eq:PM})}
        \State {\textbf{break}}
    \EndIf 
    \State {// get VN index $g$ and favored guess value $f$}
    \State {$g$, $f$ $\gets$ \textsc{SELECT-VN}($D$)}\Comment{Alg. \ref{alg:select_vn}}
    \If{\textbf{is} main branch}
    \State {$vn\_status[g]\gets\hbe[g]
    \gets$ $f$}
    \ElsIf {\textbf{is} side branch}
    \If {$D\neq D_{splitt}$}
    \State {$vn\_status[g]\gets\hbe[g]
    \gets$ $f$}
    \Else
    \State {$vn\_status[g]\gets\hbe[g]
    \gets$ $1-f$}
    \EndIf
    \Else \Comment{tree branch}
    \If {$D>4$}
    \State {$vn\_status[g]\gets\hbe[g]
    \gets$ $f$}
    \Else
    \State{// decimate accord. to the binary repr. of id}
    \State{$b\gets$ the $D^{th}$ bit of id $\in[2,15]$}
    \State {$vn\_status[g]\gets\hbe[g]\gets b+(-1)^b f$}  
    \EndIf
    
    \EndIf
\EndFor
\EndFor

\State \Return {the $\hbe$ with the smallest path metric}
\end{algorithmic}
\end{algorithm}

Let us summarize the differences between GDG and BPGD \cite{yao2023BPGD} decoder. 
First, we change the VN selection rule and make decisions based on LLR history. 
Second, we allow guessing on the main decision branch and at very early steps. 
The guessing tree has a similar spirit to guessing decoding on classical erasure channel \cite{erasure_guessing,maxwell} and later AWGN channel \cite{gaussian_guessing}. However, to our knowledge, no prior work has proposed the side branches, which is the key to convergence for our VN selection rule. It is probably because prior works decimate the most reliable VNs and guessing is unnecessary.

\subsection{Simulation}

In Fig. \ref{fig:LER}, we show the performance of using the (3,1) sliding window to decode the BB codes on circuit-level noise. 
All BP runs involved use min-sum update, flooding scheduling, and scaling factor $\alpha=1.0$. 
As described at the end of the previous section, for each window we first run BP on the original PCM for eight or sixteen iterations if $N\leq 144$ or $N=288$, rank columns according to the sum of LLR posteriors from the latest four iterations, and apply GDG to the first  $2w'$ columns. 

For $N\leq 144$, the decision tree with parameters shown in Fig. \ref{fig:decision_tree} is used as the inner decoder. The GDG critical path is $150=25\cdot 6$ BP iterations. The worst-case runtime of each window is measured to be $\sim 3ms$ for $N=144$ with a multi-thread implementation on an Intel i9-13900K CPU, while 200 BP iterations + OSD-CS10 on the original PCM takes $\sim 20ms$ or $\sim 10ms$ on the shortened PCM in the worst-case. Note that runtime on CPU is not indicative of performance on specialized hardware. Since we use BP with flooding scheduling, GDG is expected to benefit from the inherent parallel CN/VN updates on FPGA or GPU, though the VN selections (\emph{argmin} operations) may become the bottleneck. Similarly, the $\text{rank}(\bH)$ pivot findings\footnote{This argument is for higher-order OSD. For OSD-0, there is no need to do full Gaussian elimination, adding in columns can stop once the syndrome is in the span of the selected ones, see Alg. 2 of \cite{panteleev2021degenerate}.} in OSD are the bottleneck operations on ASICs. In this sense, GDG may still have a runtime advantage when parallelization is considered, since the number of required \emph{argmin} operations decreases with physical error rates. 
For the $N=288$ code, a maximum depth of $40$ is used for the main branch, and the side/tree branches are allowed to run for $20$ more steps after splitting at depth $20$ or $5$. With the same decision tree parameters applied to $(5,2)$-sliding window GDG decoding for $N=144$, we get a slightly better logical error rate than the global decoding in Fig. \ref{fig:LER}, indicating order $10$ is not sufficient enough for global OSD-CS decoding. 
Our global OSD-CS decoding curves match  those in Fig. 3 from \cite{bravyi2023highthreshold} quite well, except for the $N=288$ code. The performance degradation is possibly also due to our insufficient OSD-CS order. 
We also tried $(4,1)$-window GDG for $N=288$, it is significantly closer to the global decoding curve in Fig. \ref{fig:LER} than the one from $(3,1)$ decoding. The specific tree parameters and performance are available online.

\subsection{Do not waste guesses}
After each step of decimation, the remaining Tanner graph can have degree one CNs. The single VN connected to such a CN can be uniquely decoded as the current syndrome of this CN. This \emph{peeling} step is repeated until no degree one CN is present. Peeling adds a small overhead to the various decision paths, but it crucially prevents GDG from wasting decision steps on these apparently decidable VNs.

\begin{algorithm}
 \hspace*{\algorithmicindent} \textbf{Input} LLR history $\mathbf{\Lambda}_{hist}$ of one VN, depth $D$\\
 \hspace*{\algorithmicindent} \textbf{Output} Decimation value for this VN
\caption{Aggressive Decimation}\label{alg:aggressive_decimation}
\begin{algorithmic}[1]
\Procedure{agg-dec}{$\mathbf{\Lambda_{hist}}$, $D$}
    \State $P_A\gets -3$ \textbf{if} on main branch \textbf{else} $0$
    \State $P_B\gets -12$ \textbf{if} on main branch \textbf{else} $-10$
    \State $P_B\gets -16$ \textbf{if} $D=1$
    \State $P_C\gets 30$
    \State $P_D\gets 3$
    \If{$\mathbf{\Lambda}_{hist}<P_A$ and sum$(\mathbf{\Lambda}_{hist})<P_B$}
        \State \Return $1$
    \EndIf
    \If{($\mathbf{\Lambda}_{hist}>P_C$ \textbf{and} $D\leq 4$) \textbf{or}\\\quad($\mathbf{\Lambda}_{hist}>P_D$ and $\exists\ 
    \geq 3$ unsatisfied CN neigh.)}
        \State \Return $0$
    \EndIf
    \State {// otherwise $\mathbf{\Lambda_{hist}}$} not reliable enough for decision
    \State \Return -1
\EndProcedure
\end{algorithmic}
\end{algorithm}

As shown in Fig. \ref{fig:LER}, when used as the inner decoder of a (3,1)-sliding window decoder, the GDG decoder gives favorable performance for physical error rates $p\leq 0.002$. 
For slightly higher physical error rates, there is a higher chance that the number of single-fault mechanisms that occurred is larger than the maximum depth of the GDG. 
To improve convergence with the same maximum depth, we leave the low error mode and use Alg. \ref{alg:aggressive_decimation} aggressive decimation in Alg. \ref{alg:select_vn}.
For each VN, if its LLR history is very positive or very negative, it is simply decimated to zero or one, without performing any guessing. Line $10$ of Alg. \ref{alg:aggressive_decimation} is inspired by \cite{gaussian_guessing}, which says VNs that are connected to many unsatisfied CN neighbors are places that lack information. 
Some effort was spent on finding $P_A,P_B,P_C,P_D$ at $p=0.003$ for the $N=144$ code by first looking at posterior LLR distribution in value and then fine-tuning. The \emph{same} parameters are later applied to all other codes and $p$, without any further tuning. 
It is also important to mention that we always clip the VN to CN messages to $[-50, 50]$. The clipping values especially affect $P_C$.

We also tried Alg. \ref{alg:aggressive_decimation} in hindsight for $p\leq 0.002$. 
However, it causes slight error floors because some VNs are decimated to zero too early. 
Furthermore, the BP preprocessing iteration on the original window PCM is kept low (8 or 16), otherwise error floors appear. 
The intuition is that GDG prefers negativity in the messages, and longer BP runs that do not eventually converge ruin the negativity in the message-passing network, especially at low $p$, where most messages are positive.

\section{Conclusion}
\label{sec:conclusion}
We introduce the GDG decoder in this work and use it as the inner decoder of a sliding window decoder for BB codes under circuit-level noise. In particular, we limit the critical path of this decoder to $150$ BP iterations on each window for (3,1)-decoding of the $N=144$ code, and nevertheless GDG achieves excellent logical error rates.  However, GDG is not a general-purpose decoder like OSD. Rather, it is designed for the sub-threshold region. The average number of error weights should be smaller than the maximum step for GDG to be utilized to its full potential. For QLDPC codes, the error floor \cite{error_floor} is a conundrum, and GDG awaits special hardware implementation to investigate its error floor behavior.

Furthermore, we design GDG for syndrome decoding, not codeword decoding, e.g.\ as in Steane error correction \cite{SteaneEC}. 
We believe a similar VN selection rule for codeword decoding can be developed based on the history of $\text{sign}(\Lambda_i)\cdot \Lambda_i^{(t)}$.
Lastly, though we draw an analogy to classical SC-LDPC window decoding, we did not exploit their key feature of reusing BP messages from the previous window in the overlapping region to initialize the next window. This would be possible for the GDG decoder, using the BP internal messages from the branch with the minimum path metric, whereas no soft information can be retained for the next window when using BP+OSD. We leave such investigations for future work.

\section*{Acknowledgment}
We thank Pavel Panteleev for the idea of adding a little bit of globalness to a local decoder like BP, leading us to explore BPGD in the first place. The global operation turns out to be selecting the most likely to flip variable node for us. We also thank Joshua Viszlai for useful discussions. 
\bibliographystyle{IEEEtran}
\bibliography{main}
\appendices
\section{Data qubit noise}
\label{sec:data_qubit_noise}
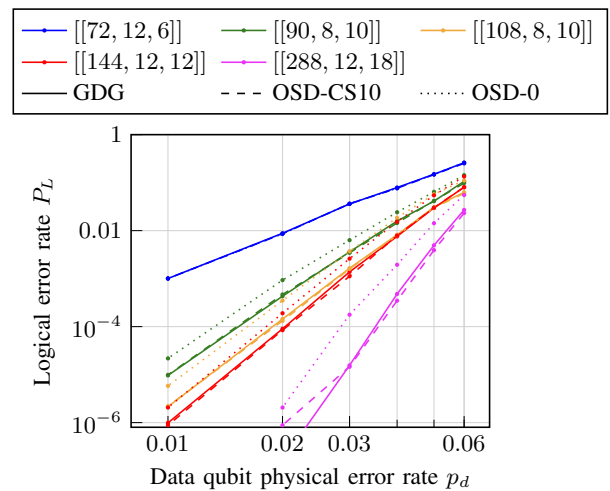
\begin{figure}[htb]
\centering
\begin{tikzpicture}
{\small
\definecolor{neonpink}{RGB}{234,51,247}
\definecolor{darkgray176}{RGB}{176,176,176}
\definecolor{lightgray211}{RGB}{211,211,211}
\definecolor{orange}{RGB}{242,169,59}
\definecolor{olivegreen}{RGB}{55,126,34}

\begin{axis}[
width=0.35\textwidth,
log basis y={10},
tick align=inside,
tick pos=left,
x grid style={lightgray211},
xlabel={\small Data qubit physical error rate \(\displaystyle p_d\)},
xmajorgrids,
xmode=log,
xmin=0.008, xmax=0.07,
xtick={0.01,0.02,0.03,0.06},
xticklabels={
  \(\displaystyle {0.01}\),
  \(\displaystyle {0.02}\),
  \(\displaystyle {0.03}\),
  \(\displaystyle {0.06}\),
},
minor xtick={0.03,0.04,0.05,0.06},
xminorgrids=true,
scaled x ticks=false,
xtick style={color=black},
y grid style={lightgray211},
ylabel={\small Logical error rate \(\displaystyle P_L\)},
ymajorgrids,
ymin=8e-7, ymax=1,
ymode=log,
semithick,
ytick style={color=black},
ytick={1e-6,0.0001,0.01,1},
yticklabels={
  \(\displaystyle {10^{-6}}\),
  \(\displaystyle {10^{-4}}\),
  \(\displaystyle {0.01}\),
  \(\displaystyle {1}\),
},
error bars/y dir=both,
error bars/y explicit,
legend cell align=left,
legend columns=3,
legend style={at={(0.5,1.25)}, /tikz/every even column/.append style={column sep=0.1cm}, fill=none, draw=black, anchor=center, align=center},
]

\addplot [blue, mark=*, mark size=0.5, mark options={solid}]
table [x=P, y=WER] {n72_k12_d6_data_noise_gdg.dat};
\addlegendentry{$[[72,12,6]]$}

\addplot [olivegreen, mark=*, mark size=0.5, mark options={solid}]
table [x=P, y=WER] {n90_k8_d10_data_noise_gdg.dat};
\addlegendentry{$[[90,8,10]]$}

\addplot [orange, mark=*, mark size=0.5, mark options={solid}]
table [x=P, y=WER] {n108_k8_d10_data_noise_gdg.dat};
\addlegendentry{$[[108,8,10]]$}

\addplot [red, mark=*, mark size=0.5, mark options={solid}]
table [x=P, y=WER] {n144_k12_d12_data_noise_gdg.dat};
\addlegendentry{$[[144,12,12]]$}

\addplot [neonpink, mark=*, mark size=0.5, mark options={solid}]
table [x=P, y=WER] {n288_k12_d18_data_noise_gdg.dat};
\addlegendentry{$[[288,12,18]]$}

\addlegendentry{}
\addlegendimage{empty legend}
\addlegendentry{GDG}
\addlegendimage{black, line legend}
\addlegendentry{OSD-CS10}
\addlegendimage{black, line legend, dashed}
\addlegendentry{OSD-0}
\addlegendimage{black, line legend, dotted}

\addplot [blue, dashed, mark=*, mark size=0.5, mark options={solid}, forget plot]
table [x=P, y=WER] {n72_k12_d6_data_noise_osd10_factor05.dat};
\addplot [olivegreen, dashed, mark=*, mark size=0.5, mark options={solid}, forget plot]
table [x=P, y=WER] {n90_k8_d10_data_noise_osd10_factor05.dat};
\addplot [orange, dashed, mark=*, mark size=0.5, mark options={solid}, forget plot]
table [x=P, y=WER] {n108_k8_d10_data_noise_osd10_factor05.dat};
\addplot [red, dashed, mark=*, mark size=0.5, mark options={solid}, forget plot]
table [x=P, y=WER] {n144_k12_d12_data_noise_osd10_factor05.dat};
\addplot [neonpink, dashed, mark=*, mark size=0.5, mark options={solid}, forget plot]
table [x=P, y=WER] {n288_k12_d18_data_noise_osd10_factor0625.dat};

\addplot [blue, dotted, mark=*, mark size=0.5, mark options={solid}, forget plot]
table [x=P, y=WER] {n72_k12_d6_data_noise_osd0_factor05.dat};
\addplot [olivegreen, dotted, mark=*, mark size=0.5, mark options={solid}, forget plot]
table [x=P, y=WER] {n90_k8_d10_data_noise_osd0_factor05.dat};
\addplot [orange, dotted, mark=*, mark size=0.5, mark options={solid}, forget plot]
table [x=P, y=WER] {n108_k8_d10_data_noise_osd0_factor05.dat};
\addplot [red, dotted, mark=*, mark size=0.5, mark options={solid}, forget plot]
table [x=P, y=WER] {n144_k12_d12_data_noise_osd0_factor05.dat};
\addplot [neonpink, dotted, mark=*, mark size=0.5, mark options={solid}, forget plot]
table [x=P, y=WER] {n288_k12_d18_data_noise_osd0_factor0625.dat};
\coordinate (right) at (rel axis cs:1,0);

\end{axis}
}
\end{tikzpicture}

\caption{Data qubit noise, X-noise only. Solid lines are GDG in low error mode with main branch maximum depth $40$, side or tree branches do not split after depth $20$ or $5$ and are allowed to proceed for $30$ more steps. Scaling factor $0.625$. Dashed and dotted lines are BP+OSD-CS10 and BP+OSD-0 respectively. BP preprocessing for both OSD methods is $100$ iterations. Scaling factor $0.5$ is used for $N\leq 144$ and $0.625$ is used for $N=288$.}
\label{fig:data_noise}
\end{figure}

In Fig. \ref{fig:data_noise}, we show the logical error rate of employing GDG or BP+OSD on data qubit noise. Here we focus on the X type of noise only and assume each data qubit is subject to i.i.d. bit flip with probability $p_d$. This equals the classical binary symmetric channel BSC($p_d$).
The scaling factor used for GDG is always set to $0.625$. For OSD methods, the best scaling factor is chosen among $\{0.5,0.625,0.8,1.0\}$.
The $\bH_X$ and $\bH_Z$ have girth $6$ for these BB codes, when decoding $X$ and $Z$ separately, though we still use a history length of $4$ for GDG as in the main text.
The three methods are run over the same set of generated noise, and for the $N=288$ at $p_d=0.02$, OSD-CS10 shows an error floor while GDG does not.

We did not perform simulation on data depolarizing channels, where each data qubit independently and
identically experiences a random $X$, $Z$, or $Y = iXZ$ type of error, each with probability $p/3$ for some $p\in [0, 1]$\footnote{As a rule of thumb, decoding on Depolarize($p$) is easier to decode than BSC($2p/3$) since the correlation between X and Z can be exploited.}, as a quaternary BP (BP4) version for GDG should be designed and compared to BP4+OSD\footnote{An implementation can be found in our repository online. Additionally, our TensorFlow implementation of BP4+OSD-0 can be found in https://github.com/gongaa/Feedback-GNN/blob/main/examples/OSD.ipynb.}.

\begin{figure}[htb]
\centering
\begin{tikzpicture}
{
\definecolor{neonpink}{RGB}{234,51,247}
\definecolor{darkgray176}{RGB}{176,176,176}
\definecolor{lightgray211}{RGB}{211,211,211}
\definecolor{orange}{RGB}{242,169,59}
\definecolor{olivegreen}{RGB}{55,126,34}
\definecolor{mulberry}{RGB}{198,75,140}
\definecolor{lilac}{RGB}{182,96,205}
\definecolor{lavender}{RGB}{228,160,247}

\begin{axis}[
width=0.3\textwidth,
log basis y={10},
tick align=inside,
tick pos=left,
x grid style={lightgray211},
xlabel={\small Syndrome flip error rate \(\displaystyle p_s\)},
xmajorgrids,
xmode=log,
xmin=1e-5, xmax=1e-3,
xtick={1e-5,1e-4,1e-3},
xticklabels={
  \(\displaystyle {10^{-5}}\),
  \(\displaystyle {10^{-4}}\),
  \(\displaystyle {0.001}\),
},
minor xtick={2e-5,5e-5,2e-4,5e-4},
xminorgrids=true,
scaled x ticks=false,
xtick style={color=black},
y grid style={lightgray211},
ylabel={\small Logical error rate \(\displaystyle P_L\)},
ymajorgrids,
ymin=1e-5, ymax=0.1,
ymode=log,
semithick,
ytick style={color=black},
ytick={1e-5,1e-4,0.001,0.01,0.1},
yticklabels={
  \(\displaystyle {10^{-5}}\),
  \(\displaystyle {10^{-4}}\),
  \(\displaystyle {0.001}\),
  \(\displaystyle {0.01}\),
  \(\displaystyle {0.1}\),
},
error bars/y dir=both,
error bars/y explicit,
transpose legend,
legend style={at={(1.2,1)},anchor=north west, font=\small, text opacity=1, draw=none},
legend cell align=left,
]

\addlegendentry{\textbf{Method}}
\addlegendimage{empty legend}
\addlegendentry{GDG}
\addlegendimage{black, line legend}
\addlegendentry{OSD-CS10}
\addlegendimage{black, line legend, dashed}
\addlegendentry{lower bound}
\addlegendimage{black, line legend, dotted}
\addlegendentry{\textbf{Data noise}}
\addlegendimage{empty legend}

\addplot [neonpink, mark=*, mark size=0.5, mark options={solid}]
table [x=Psynd, y=WER] {n288_p06_phe_gdg.dat};
\addlegendentry{$p_d=0.06$}

\addplot [neonpink, dashed, mark=*, mark size=0.5, mark options={solid}, forget plot]
table [x=Psynd, y=WER] {n288_p06_phe_osd10.dat};

\addplot [lilac, mark=diamond*, mark size=1.0, mark options={solid}]
table [x=Psynd, y=WER] {n288_p05_phe_gdg.dat};
\addlegendentry{$p_d=0.05$}

\addplot [lilac, dashed, mark=diamond*, mark size=1.0, mark options={solid}, forget plot]
table [x=Psynd, y=WER] {n288_p05_phe_osd10.dat};

\addplot [lavender, mark=square, mark size=1.0, mark options={solid}]
table [x=Psynd, y=WER] {n288_p04_phe_gdg.dat};
\addlegendentry{$p_d=0.04$}

\addplot [lavender, dashed, mark=square, mark size=1.0, mark options={solid}, forget plot]
table [x=Psynd, y=WER] {n288_p04_phe_osd10.dat};

\addplot[lavender, dotted, samples=100, domain=1e-5:1e-3, forget plot] {3.47e-4+864*x^2+864*0.04*3*x^2};

\addplot [mulberry, mark=triangle, mark size=1.0, mark options={solid}]
table [x=Psynd, y=WER] {n288_p03_phe_gdg.dat};
\addlegendentry{$p_d=0.03$}

\addplot [mulberry, dashed, mark=triangle, mark size=1.0, mark options={solid}, forget plot]
table [x=Psynd, y=WER] {n288_p03_phe_osd10.dat};

\addplot[mulberry, dotted, samples=100, domain=1e-5:1e-3, forget plot] {1.46e-5+864*x^2+864*0.03*3*x^2};


\end{axis}
}
\end{tikzpicture}
\vspace*{-4mm}
\caption{Logical error rate for the $\llbracket 288,12,18\rrbracket$ code at data qubit X-noise $p_d=0.03 \sim 0.06$ and i.i.d. syndrome bit-flip with probability $p_s\in [10^{-5}, 10^{-3}]$. The dotted lines are the lower bound $p_L(p_d)+864\cdot p_s^2+2592\cdot p_d\cdot p_s^2$.}
\label{fig:phenomenonlogical}
\end{figure}
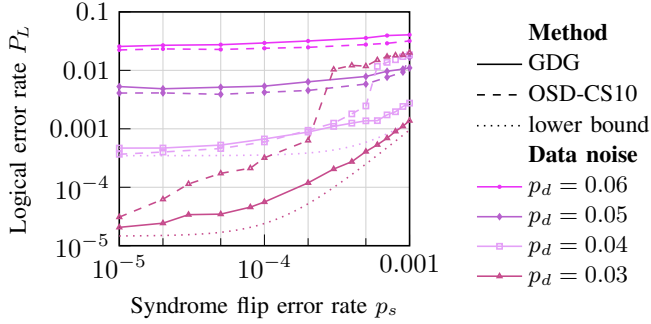

\section{Single shot syndrome noise}
\label{sec:single_shot}
In the main text, so far we have been investigating using BB codes as a fault-tolerant memory. The simulation we performed is also an emulation of the memory experiment. The final round of the noiseless syndrome measurement can be obtained by measuring (collapsing) all the data qubits at the end and followed by \emph{classical} syndrome calculation. However, when an error correction code is used in a real-time experiment, it is not clear how to obtain noiseless syndrome, since the data qubits shall not be collapsed before a non-Clifford gate and we have to deal with measurement noise.

The lift-product construction \cite{almost_linear_distance} creates a naturally overcomplete PCM. The GHP codes in \cite{panteleev2021degenerate}, 2BGA codes \cite{2BGA} and the BB codes \cite{bravyi2023highthreshold} are small special cases. In this section, we investigate the syndrome protection offered by the syndrome code in the single-shot noisy syndrome measurement setting for the $\llbracket 288,12,18\rrbracket$ BB code.

In Fig. \ref{fig:phenomenonlogical}, assume the data qubit noise is subject to $\be\sim$ BSC($p_d$) and the syndrome noise is subject to $\bs'\sim$ BSC($p_s$). Throughout we assume $p_s<p_d$, otherwise it makes no sense to perform measurements. The observed syndrome is thus $\bH_X\be + \bs'$ and the decoding $\hbe$ is deemed successful if $\bH_Z^{\perp}(\hbe+\be)=\mathbf{0}$. The row span of $\bH_Z^{\perp}$ contains $\bH_X$ by the CSS constraint, and thus $\hbe$ needs to have the same syndrome as $\be$ for the decoding to be correct. 
In Tanner graph construction, we add one virtual VN to each CN, the flipping of such a virtual VN implies a syndrome flip on the associated CN. The effective PCM is then $[\bH_X|\mathbf{I}]$. The prior LLR to the virtual and original VNs are $\log \frac{1-p_s}{p_s}$ and $\log \frac{1-p_d}{p_d}$, respectively. For the $\llbracket 288,12,18\rrbracket$ code with the PCMs specified in \cite{bravyi2023highthreshold}, there are $288$ original VNs of degree three and $144$ virtual VNs of degree one. We employ BP+OSD-CS10 and GDG to decode $\hbe$. The same parameters as in Fig. \ref{fig:data_noise} are used for both decoders. 

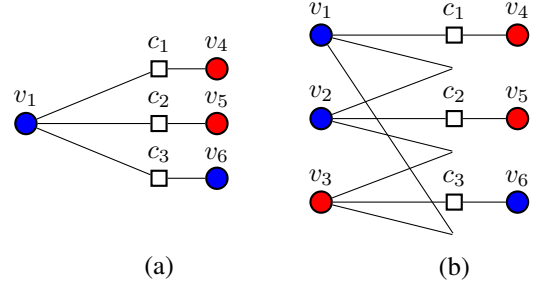
\begin{figure}
    \centering

\usetikzlibrary{shapes.geometric}
\tikzstyle{cnd} = [rectangle,draw=black,fill=white,thick,text width=2mm,text height=2mm,inner sep=0pt]
\tikzstyle{cndvoid} = [draw=white,fill=white,inner sep=0pt]

\tikzstyle{vnd} = [circle,draw=black,fill=white,thick,text width=2mm,inner sep=0pt]
\tikzstyle{vnd1} = [circle,draw=black,fill=red,thick,text width=2mm,text height=2mm,inner sep=0pt]
\tikzstyle{vnd2} = [circle,draw=black,fill=blue,thick,text width=2mm,text height=2mm,inner sep=0pt]

\begin{tikzpicture}[scale=0.33, baseline={(0,-2.2)}]
\label{fig:wt1}
	    \node [vnd2, label={[yshift=-0.05cm,xshift=0.cm]$v_1$}] (vnd1) {};
   
	    \node [cnd, right=1.5cm of vnd1, label={[yshift=-0.02cm,xshift=0.cm]$c_2$}] (cnd2) {};
        \node [cnd, above=0.5cm of cnd2, label={[yshift=-0.02cm,xshift=0.cm]$c_1$}] (cnd1) {};   
	    \node [cnd, below=0.5cm of cnd2, label={[yshift=-0.02cm,xshift=0.cm]$c_3$}] (cnd3) {};

        \node [below=0.8cm of cnd3] {(a)};
	    \node [vnd1, right=0.5cm of cnd1,label={[yshift=-0.05cm,xshift=0.cm]$v_4$}] (vnd4) {};
	    \node [vnd1, right=0.5cm of cnd2, label={[yshift=-0.06cm,xshift=-0.cm]$v_5$}] (vnd5) {};
	    \node [vnd2, right=0.5cm of cnd3, label={[yshift=-0.07cm,xshift=-0.cm]$v_6$}] (vnd6) {};

        \path[-,black] (vnd1) edge node [] {} (cnd1);
        \path[-,black] (vnd1) edge node [] {} (cnd2);
        \path[-,black] (vnd1) edge node [] {} (cnd3);
        
        \path[-,black] (cnd1) edge node [] {} (vnd4);
        \path[-,black] (cnd2) edge node [] {} (vnd5);
        \path[-,black] (cnd3) edge node [] {} (vnd6);

\end{tikzpicture}
\qquad
\begin{tikzpicture}[scale=0.33]
\label{fig:wt3}
	    \node [vnd2, label={[yshift=-0.05cm,xshift=0.cm]$v_1$}] (vnd1) {};
	    \node [vnd2, below=0.8cm of vnd1, label={[yshift=-0.05cm,xshift=-0.cm]$v_2$}] (vnd2) {};
	    \node [vnd1, below=0.8cm of vnd2, label={[yshift=-0.05cm,xshift=-0.cm]$v_3$}] (vnd3) {};
              
	    \node [cnd, right=1.5cm of vnd1, label={[yshift=-0.02cm,xshift=0.cm]$c_1$}] (cnd1) {};
	    \node [cndvoid, below=0.3cm of cnd1] (cnd4) {};     
	    \node [cnd, right=1.5cm of vnd2, label={[yshift=-0.02cm,xshift=0.cm]$c_2$}] (cnd2) {};
	    \node [cndvoid, below=0.3cm of cnd2] (cnd5) {};
	    \node [cnd, right=1.5cm of vnd3, label={[yshift=-0.02cm,xshift=0.cm]$c_3$}] (cnd3) {};
	    \node [cndvoid, below=0.3cm of cnd3] (cnd6) {};

        \node [below=0.15cm of cnd6] {(b)};

	    \node [vnd1, right=2.3cm of vnd1,label={[yshift=-0.05cm,xshift=0.cm]$v_4$}] (vnd4) {};
	    \node [vnd1, right=2.3cm of vnd2, label={[yshift=-0.06cm,xshift=-0.cm]$v_5$}] (vnd5) {};
	    \node [vnd2, right=2.3cm of vnd3, label={[yshift=-0.07cm,xshift=-0.cm]$v_6$}] (vnd6) {};

        \path[-,black] (vnd1) edge node [] {} (cnd1);
        \path[-,black] (vnd2) edge node [] {} (cnd2);
        \path[-,black] (vnd3) edge node [] {} (cnd3);

        \path[-,black] (vnd1) edge node [] {} (cnd4);
        \path[-,black] (vnd1) edge node [] {} (cnd6);
        \path[-,black] (vnd2) edge node [] {} (cnd4);
        \path[-,black] (vnd2) edge node [] {} (cnd5);
        \path[-,black] (vnd3) edge node [] {} (cnd5);
        \path[-,black] (vnd3) edge node [] {} (cnd6);
        
        \path[-,black] (cnd1) edge node [] {} (vnd4);
        \path[-,black] (cnd2) edge node [] {} (vnd5);
        \path[-,black] (cnd3) edge node [] {} (vnd6);

\end{tikzpicture}
    \vspace*{-4mm}
    \caption{Tanner graph of the data and syndrome noise decoding, only showing relevant VNs/CNs. The VNs on the left of CNs are data qubits, and the VNs on the right of CNs are virtual nodes for syndrome noise. The red and blue configurations cause the same syndrome. If the red configuration happens, a maximum likelihood decoder will choose the more probable blue one because $p_d > p_s$ and get the denoised syndrome wrong, introducing logical errors.}
    \label{fig:tanner_graph}
\end{figure}

Next, we discuss how to obtain the lower bound for the logical error rate in Fig. \ref{fig:phenomenonlogical}. 
The first term $p_L(p_d)$ is the logical error rate in the absence of syndrome error\footnote{Here we ignore wrong decoding with a wrong denoised syndrome but accidentally lead to the correct answer.}, which we obtain from Fig. \ref{fig:data_noise}. The second term $864\cdot p_s^2=\binom{3}{2}\cdot 288\cdot p_s^2$ comes from the configurations in Fig. \ref{fig:tanner_graph}(a) where weight-two syndrome error happens on the three CNs neighbors of an original VN. The third term $2592\cdot p_d\cdot p_s^2$ originates from a weight-one data qubit error and weight-two syndrome error in configuration Fig. \ref{fig:tanner_graph}(b). This term requires careful counting for the coefficient.

Consider the $\bH_X=[\bA|\bB]$ matrix of the $N=288$ BB code, where $\bA=x^3+y^2+y^7$, $\bB=y^3+x+x^2$ and $x=\mathbf{S}_{12}\otimes \mathbf{I}_{12}$, $y=\mathbf{I}_{12}\otimes \mathbf{S}_{12}$, $\mathbf{S}_{12}$ is the cyclic permuting matrix. 
The syndrome code for $\bH_X$ is its column span. Its shape $144\times 288$ implies that each column is a codeword of length $144$ and the syndrome code is spanned by its $288$ linearly dependent columns, each of weight-three. The syndrome code can be written as the ideal $(b_1, b_2)$ in the quotient ring $R=\F_2[x,y]/(x^{12}=1,y^{12}=1)$. The two base polynomials  $b_1=y^5+y^{10}+x^9$ and $b_2=y^9+x^{10}+x^{11}$ are the first columns of matrix $A$ and $B$ respectively. A monomial $x^a y^b$ present in the polynomial indicates a one at position $12a+b$. Then the $288$ columns of $\bH_X$ are just $b_1\cdot x^a y^b$ and $b_2\cdot x^a y^b$, $0\leq a,b\leq 11$, and it is easy to see they all have weight three. Since the syndrome code is a linear code, the distance is upper bound by three. Indeed, the column span of $\bA$ and $\bB$ each has distance three, while the column span of $\bH_X=[\bA|\bB]$ has distance two. Importantly, this does not imply weight one syndrome errors will cause decoding failure. In fact, to denoise a weight-one syndrome noise to a weight-two syndrome codeword, an additional 14 VNs need to be flipped as well. For the number of configurations in Fig. \ref{fig:tanner_graph}(b), we need to count the number of weight-three syndrome codewords caused by three original VNs. They are $b_1^2=y^{10}+y^{20}+x^{18}=y^5\cdot b_1+y^{10}\cdot b_1+x^{9}\cdot b_1$ or $b_2^2$ and their $144$ shifts by $x^a y^b$. Therefore, all together there are $\binom{3}{2}\cdot\binom{3}{1}\cdot 2\cdot 144=2592$ such configurations.

The above arguments hint that a higher level of syndrome protection may be achieved by a larger distance of the syndrome code, which is upper bound by the column weight of the PCM. The $\llbracket 254,28,14\leq d\leq 20\rrbracket$ code in \cite{panteleev2021degenerate} is also a bicycle code and the generator polynomials for $\bA$ and $\bB$ are $a(x)=1+x^{15}+x^{20}+x^{28}+x^{66}$ and $b(x)=1+x^{58}+x^{59}+x^{100}+x^{121}$, where $x$ is the cyclic permuting matrix of size $127$. Its syndrome code has distance five since $g(x)\ \vert\  \text{gcd}(a(x),b(x))$ and $g(x)=g_1(x)\cdot g_2(x)=(x^7+x+1)\cdot(x^7+x^5+x^3+x+1)$ is the generator polynomial of the $[127,14,5]$ BCH code\footnote{The distance lower bound follows from the BCH bound. Assume $g_1(\alpha)=0$, then $\alpha^2,\alpha^4$ are also roots. Further, since $g_1(\alpha)\ \vert\  g_2(\alpha^3)$, $\alpha^3$ is also a root of $g(\alpha)$.}. The first dominant term to the logical error rate can be obtained similarly to Fig. \ref{fig:tanner_graph}(a) and is $254\cdot \binom{5}{3}\cdot p_s^3$. Simulation with GDG shows that this code can achieve $\sim 10^{-7}$ logical error rate at $p_d=0.01$ and $p_s=10^{-4}$. On the contrary, the logical error rate for the $\llbracket 288,12,18 \rrbracket$ code at $p_s=10^{-4}$ is at least $864\cdot p_s^2\approx 10^{-5}$.

It should be mentioned that GDG can handle soft syndrome information as well, instead of using $\log \frac{1-p_s}{p_s}$ as the prior LLR for the virtual VNs in BSC($p_s$), an input-dependent LLR can be used, see Eq. (1) of \cite{raveendran2022soft} for Gaussian syndrome noise.
Moreover, in Fig. 4(a) of \cite{raveendran2022soft}, a phase transition was observed when sweeping the standard deviation of the Gaussian noise, despite using a different decoding method on a different code.

\section{GDG implementation}
We use the \texttt{mod2sparse} library for sparse matrix manipulation, the same library was used by the BP+OSD package \cite{Roffe_LDPC_Python_tools_2022} by Roffe et al. To handle decimations, we additionally maintain a status snapshot containing ternary masks for VNs (decided with 0/1, undecided) and CNs (active with 0/1, resolved).
In message passing, CN/VN updates simply ignore messages coming from inactive neighboring VNs/CNs.

To facilitate intermediate peeling steps, we also maintain a vector for CN degrees, which is updated whenever VN decimation is performed and hence is a part of the status snapshot. Peeling stops when all CN degrees are larger than one. It is possible that \emph{contradictions} happen during peeling, e.g., two degree-one CNs want to set contradicting values for their common VN neighbor. In this case, the current decision path is immediately killed and it returns PM $\infty$. 

There is also a vector for VN degrees so that degree one or two VNs can be easily skipped in Alg. \ref{alg:select_vn}. This vector never needs to be updated, since decimation, no matter induced by VN selection or peeling, does not cause the degree of \emph{active} VN to change.

In Fig. \ref{fig:decision_tree}, since previous decisions affect the later, \emph{different decision paths at the same depth may not decimate the same VN}. Moreover, BP iterations are reused as much as possible. For example, the two neighboring tree branches share the first few decisions, and each side branch has some overlap with the main branch. In this case, at the splitting step, the status snapshot is saved and the selected VN is decimated to the favored value first. After this direction is finished, the status snapshot is reloaded and decimation in the other direction is performed. As only the CN/VN status is saved but not the BP messages for the unfavored direction, the message-passing network has to be re-initialized.

We target an Intel i9-13900K CPU that has $32$ logical cores during development. For the multi-thread implementation, apart from the main thread associated to the main branch, we have tree threads that each handle two neighboring tree branches, and side branches each handling a side branch. Each thread is set affinity to one CPU logical core. This assignment enables us to use $32$ threads for the $N=288$ code. Tree threads run independently from the main thread after copying the column-permuted PCM, starting from the root of the decision tree. Side threads copy columns and wait until the main thread writes to it the status snapshot. It is also possible to decouple the side threads from the main thread. They can run from the root and compute the overlapping decision paths themselves due to the deterministic nature of guided decimation. By this means, inter-core communication can also be minimized. However, we did not follow this approach due to the P-core and E-core differences in this Intel CPU. Side threads are assigned to the slower E-cores for the $N=288$ codes.

The single-thread version is more straightforward. It first proceeds with the main branch and saves the snapshots along the way. After completion of the main branch, it reloads each snapshot and runs BP, again saving the snapshot if guessing is still allowed. It terminates when all the snapshots are consumed. Since the BP messages are not retained by reloading snapshots, the single-thread version re-initializes BP more often in the guessing tree region. Moreover, some early-stop heuristics are implemented for the single-thread version, improving amortized runtime but causing further differences to the multi-thread version.

\section{Stim circuit implementation}
Our Stim \cite{stim} circuit implementation nearly follows the surface code memory experiments. When doing the memory experiment in the $Z$ or $X$ basis, all the data qubits are initialized in $|0\ra$ or $|+\ra$ state. If $R$ rounds are specified by the user, including the noisy encoding round, the noisy SM circuit is repeated $R$ times. Finally, a noiseless round of SM is added to capture the final error. We implement this step by measuring all the data qubits in the $Z$ or $X$ basis and doing classical syndrome calculations afterward. Different from the surface code implementation, we do not add flips before this final data qubit measurement. Another difference is that we always use \texttt{RX} or \texttt{MRX}, instead of applying noisy Hardamard gates before and after \texttt{R} or \texttt{MR} gates, as it is the practice in \cite{high_threshold_fowler_2009}. The above differences in circuit creation do not cause changes in the parity-check matrix of the circuit code when $X$ and $Z$ are decoded separately. They only affect the prior probabilities, and GDG is robust to these changes. The logical error rate per round $p_L$ is calculated from the total logical error rate $P_{L,R}$ as $p_L=1-(1-P_{L,R})^{1/R}$.

It is important to mention that we always require both Eq. (\ref{eq:overall}) and Eq. (\ref{eq:logical}) to hold for the decoding to be successful. This stricter requirement may lead to a slight overestimation of the logical error rate for GDG due to potential syndrome inconsistency in Eq. (\ref{eq:overall}). OSD remains unaffected because it has no consistency issue thanks to Gaussian elimination. To see why overestimation can happen for GDG, assume GDG fails Eq. (\ref{eq:overall}) because its estimated fault pattern and the actual one differ \emph{only} by some measurement faults. Measurement faults have no end effect, that is to say, the estimated fault propagates to the same final error string as the actual fault, which of course introduces no logical error. Note that in this case, the estimated fault should satisfy Eq. (\ref{eq:logical}). 

However, we believe the overestimation of GDG logical error is negligible, because it is very rare (less than one percent of the total failure cases) that GDG gives an estimation that fails Eq. (\ref{eq:overall}) but satisfies Eq. (\ref{eq:logical}). Even when it happens, OSD-CS10 usually induces a logical error when decoding them, therefore we tend to believe these rare cases are inherently not decodable. 

A more formal definition of the two logical error criteria and the relation between them is as follows. Consider a memory experiment in the Z basis. Call the actual final X-type error string $\bx$, which is captured by the noiseless Z-type checks $\bH_Z$ added to the end. The estimated final X-type error $\hbx$ does not introduce a logical error if and only if $\hbx$ and $\bx$ differ by an X-type stabilizer. This criterion can be written as $\bH_X^{\perp}(\bx+\hbx)=\mathbf{0}$. The row span of $\bH_X^{\perp}$ contains $\bH_Z$, and $\bH_X^{\perp}\backslash \bH_Z$ is the logical Z operator $\bL_Z$. Therefore, both $\bH_Z(\bx+\hbx)=\mathbf{0}$ and $\bL_Z(\bx+\hbx)=\mathbf{0}$ need to be satisfied. The second equation is just a rephrased version of Eq. (\ref{eq:logical}), but Eq. (\ref{eq:overall}) strictly implies the first equation.


Another subtle thing is that, in Fig. \ref{fig:visualization}, the top left $\bH_0$ is actually $\bH'_0$ due to the encoding round difference. The columns of $\bH'_0$ are contained in $\bH_0$ and $\bH'_0$ has shape $w\times 2w$. The $2w$ columns come from fault equivalently happening individually on L- or R-data before all CNOT gates in Fig. 7 of \cite{bravyi2023highthreshold}. In later SM round, $\bH_0$ has $w$ more columns than $\bH'_0$, they originate from the fault happening just before the block CNOT gates in round $6$ on X-checks in Fig. 7 of \cite{bravyi2023highthreshold}. One can see this kind of fault propagates to two faults on L-data qubits before the next cycle starts, and cannot be caught by Z-check measurements in the current cycle.

\section{Miscellaneous discussions}
\label{sec:misc}
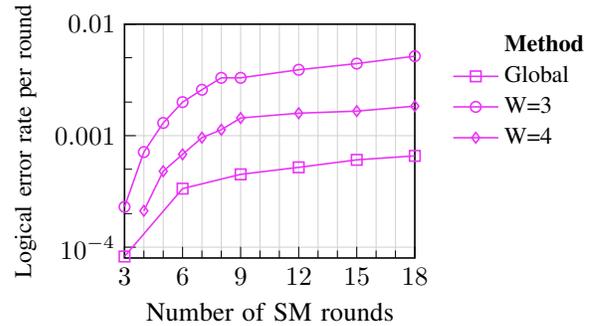
\begin{figure}[htb]
\centering
\begin{tikzpicture}
{
\definecolor{neonpink}{RGB}{234,51,247}
\definecolor{darkgray176}{RGB}{176,176,176}
\definecolor{lightgray211}{RGB}{211,211,211}
\definecolor{orange}{RGB}{242,169,59}
\definecolor{olivegreen}{RGB}{55,126,34}


\begin{axis}[
width=0.3\textwidth,
log basis y={10},
tick align=inside,
tick pos=left,
x grid style={lightgray211},
xlabel={Number of SM rounds},
xmajorgrids,
xmin=3, xmax=18,
xtick={3,6,9,12,15,18},
minor xtick={4,5,7,8,10,11,13,14,16,17},
xminorgrids=true,
scaled x ticks=false,
xtick style={color=black},
y grid style={lightgray211},
ylabel={\small Logical error rate per round},
ymajorgrids,
ymin=8e-5, ymax=0.01,
ymode=log,
semithick,
ytick style={color=black},
ytick={1e-4,0.001,0.01},
yticklabels={
  \(\displaystyle {10^{-4}}\),
  \(\displaystyle {0.001}\),
  \(\displaystyle {0.01}\),
},
minor ytick={2e-4,5e-4,2e-3,5e-3},
error bars/y dir=both,
error bars/y explicit,
transpose legend,
legend style={at={(1.1,1)},anchor=north west, font=\small, text opacity=1, draw=none},
legend cell align=left,
]

\addlegendentry{\textbf{Method}}
\addlegendimage{empty legend}

\addplot [neonpink, mark=square, mark size=2.0, mark options={solid}]
table [x=repeat, y=LER] {n288_k12_d18_overall_repeat.dat};
\addlegendentry{Global}

\addplot [neonpink, mark=o, mark size=2.0, mark options={solid}]
table [x=repeat, y=LER] {n288_k12_d18_W3_repeat.dat};
\addlegendentry{W=3}

\addplot [neonpink, mark=diamond, mark size=2.0, mark options={solid}]
table [x=repeat, y=LER] {n288_k12_d18_W4_repeat.dat};
\addlegendentry{W=4}

\end{axis}
}
\end{tikzpicture}
\vspace*{-4mm}
\caption{Logical error rate per round when changing the number of syndrome measurement rounds for the $\llbracket 288, 12,18\rrbracket$ code at physical error rate $0.005$. BP+OSD-CS10 is used for global decoding. For the (3,1) or (4,1) sliding-window decoding, GDG is used on all but the last window and BP+OSD-CS10 on the last window.}
\label{fig:window}
\end{figure}

For the $\llbracket 288,12,18\rrbracket$ code, a noticeable performance degradation between our global decoding curve in Fig. \ref{fig:LER} and Fig. 3 from \cite{bravyi2023highthreshold} is observed. This is possibly due to the insufficient OSD-CS order used by us.
Furthermore, in Fig. \ref{fig:LER}, the (3,1)-sliding window decoder also has a larger factor of performance loss compared to shorter block-length codes, since the window size $3$ is too small compared to the code distance $18$. In Fig. \ref{fig:window}, one can see that a (4,1)-sliding window decoder manages to bridge the performance gap by half in logarithmic scales. For this high physical error rate $0.005$, the GDG on the last window is replaced by BP+OSD-CS10 in both sliding window decoders, which gives roughly a factor of two performance improvement when SM is repeated for $3$ rounds. However, the improvement gradually decreases to almost no effect for $18$ rounds.

Throughout our simulation, we find that GDG outperforms BP+OSD-CS10 on all the previous windows, in terms of path metric, but cannot guarantee absolute convergence in the last window where noiseless SM is present, leading to slight performance loss. Replacing GDG with BP+OSD-CS10 in the last window usually gives a little performance boost (less than $20$ percent) for the $N\leq 144$ codes, but this boost gets smaller when more SM rounds are used. We did not try to improve GDG on the last window in this work, but it is certainly worth investigating for future work.

Another possible improvement to GDG is automorphism ensemble decoding. In Appendix \ref{sec:single_shot}, when decoding the 216 weight-two syndrome codewords, whose minimum weight solution consists of 14 VNs, GDG only succeeds in two of them. In data qubit decoding, the Tanner graph is highly symmetric and a lot of VNs have the same history, a better heuristic for guided decimation should be developed to break ties.

\end{document}